\documentclass[twocolumn,9pt]{article}

\usepackage[utf8]{inputenc}
\usepackage[T1]{fontenc}

\usepackage[a4paper, total={7in, 10in}]{geometry}
\usepackage[parfill]{parskip}
\usepackage[ruled]{algorithm2e}
\usepackage[separate-uncertainty=true]{siunitx}
\usepackage{hyperref}
\usepackage{nicefrac}
\usepackage{aas_macros}
\usepackage{natbib}
\usepackage{graphicx}
\usepackage{physics}
\usepackage{subcaption}
\usepackage{booktabs, multirow}
\usepackage{cleveref}
\usepackage{authblk}

\usepackage{enumitem} 

\DeclareSIUnit\jansky{Jy}
\DeclareSIUnit\flop{Flop}
\DeclareSIUnit\byte{B}

\DeclareCaptionFormat{custom}
{%
    \textbf{#1#2}{\small #3}
}
\newcommand\marklessfootnote[1]{
    \addtocounter{footnote}{1} 
    \footnotetext{#1}
}

\title{
That pesky A-term: Efficiently correcting for \\
direction-, time-, and baseline-dependent effects \\
in radio interferometric imaging
}
\author[1]{Torrance Hodgson}
\affil[1]{\small{Curtin Institute for Data Science, Curtin University}}
\date{January 2025}

\begin{document}


\twocolumn[
    \begin{@twocolumnfalse}
    \maketitle
    \begin{abstract}
        \qquad
        
        \noindent Radio interferometers must grapple with apparent fields of view that distort the true radio sky. These so-called `$A$-term' distortions may be direction-, time- and baseline-dependent, and include effects like the primary beam and the ionosphere. Traditionally, properly handling these effects has been computationally expensive and, instead, less accurate, ad-hoc methods have been employed. Image domain gridding (\textsc{idg}; van der Tol et al., 2018) is a recently developed algorithm that promises to account for these $A$-terms both accurately and efficiently. Here we describe a new implementation of \textsc{idg} known as the Parallel Interferometric \textsc{gpu} Imager (Pigi). Pigi is capable of imaging at rates of almost half a billion visibilities per second on modest hardware, making it well suited for the projected data rates of the Square Kilometre Array, and is compatible with both \textsc{nvidia} and \textsc{amd} \textsc{gpu} hardware. Its accuracy is principally limited only by the degree to which $A$-terms are spatially sampled. Using simulated data from the Murchison Widefield Array, we demonstrate the extraordinary effectiveness of Pigi in correcting for extreme ionospheric effects, and point to future work that would enable these results on real-world data.
    \end{abstract}
    \qquad
    \end{@twocolumnfalse}
]

\section{Introduction}

\marklessfootnote{Correspondence: \href{mailto:torrance@pravic.xyz}{torrance@pravic.xyz}}Radio interferometers must grapple with apparent fields of view that distort the true radio sky. These distortions may be direction-, time- and baseline-dependent, and include effects like the primary beam and ionospheric phase delays. In the radio interferometer measurement equation \citep{Hamaker2000, Smirnov2011}, these effects are collapsed in the order they are encountered into a single term, $A(l, m, t)$, giving the moniker `$A$-term' effects.

The discrete imaging equation can account for the $A$-term but is prohibitively expensive to compute directly. It scales with both pixel and visibility count, and involves the computation of expensive trigonometric functions. More recent interferometer designs have pushed this computational complexity ever higher. Arrays now typically include hundreds of stations, causing a combinatorial explosion of baselines and therefore visibility data, and meanwhile image sizes have also grown, necessitated by the combination of wide fields of view as well as increasingly longer baselines.

Most imaging algorithms have mitigated these computational costs by instead using some form of convolutional gridding that in turn allows for the fast Fourier transform (\textsc{fft}) to efficiently shift between the visibility and image domains. Convolutional gridding takes the sparse and irregular visibility samples of a real-world interferometer and interpolates them onto a grid by a combination of convolution and resampling operations \citep{Brouw2975}. Traditionally, the $A$-term was ignored altogether, and by imaging only narrow fields of view the so-called $w$-term could similarly be ignored, reducing the problem to a simple two-dimensional transform. Over time, however, the evolution of interferometer design and increasing fidelity requirements have necessitated folding in these terms as additional convolution kernels. W-projection \citep{Cornwell2008}, for example, addresses the $w-$term by computing Fresnel kernels for each $w$ coordinate, and AW-projection \citep{Bhatnagar2008} folds in further kernels to handle beam and ionospheric effects that may need to be recomputed per baseline and over short time intervals. To maintain accuracy, these kernels must be highly oversampled \citep[e.g.][]{Offringa2019}, and the computation of these kernels can quickly dominate the algorithm complexity costs.

In practice, due to these complexity costs modern radio imaging workflows have instead opted to handle the $A$-term using simpler alternatives that trade computational costs for accuracy. The Murchison Widefield Array \citep[\textsc{mwa};][]{Tingay2013}, for example, uses an imaging pipeline that employs short-duration `snapshot' imaging and image-based stacking \citep[e.g.][]{Wayth2015}. In this technique, $A$-terms are assumed constant over the snapshot duration and corrections are approximated in the image domain. Meanwhile, the Low Frequency Array \citep[\textsc{lofar};][]{vanHaarlem2013} most commonly uses some form of faceting \citep[e.g.][]{Williams2021}, where the field of view is decomposed into multiple small facets. Each facet is small enough that direction-dependent effects can be considered negligible, and time- and baseline-dependent effects are corrected as a global gains correction.

The image domain gridding \citep[\textsc{idg};][]{VanDerTol2018} algorithm is a promising alternative to AW-projection that can accurately apply time-, direction- and baseline-dependent $A$-terms as part of the (de)gridding operation. As the name suggests, it grids in the image rather than visibility domain where various $A$-term functions typically have a more natural expression. Critically, its core (de)gridding algorithms are embarrassingly parallel and can be implemented as \textsc{gpu} kernels. \textsc{idg} has been implemented by its authors, showing good performance on both \textsc{cpu} and \textsc{nvidia} \textsc{gpu} hardware \citep{Veenboer2017}, and it has been shown capable of keeping pace with the projected data rates of the upcoming Square Kilometre Array (\textsc{ska}) Low with only modest hardware requirements \citep{Veenboer2020}.

In this paper, we present the Parallel Interferometric \textsc{gpu} Imager (Pigi), a new implementation of the \textsc{idg} algorithm. Pigi was written with the aim to ensure compatibility with both \textsc{nvidia} and \textsc{amd} \textsc{gpu} hardware, with the latter recently being used in large supercomputer installations such as Frontier \citep{Frontier} and Setonix \citep{Pawsey}. In this paper, we provide important implementation details of Pigi, show that the algorithm achieves high levels of accuracy, and provide detailed performance benchmarks that ensure that Pigi is well suited to the demands of real-world interferometry in the \textsc{ska} era. Finally, we explore some early case studies of $A$-term corrections with the \textsc{mwa}. To begin, however, let us review the \textsc{idg} algorithm itself.

\section{Image domain gridding}

\textsc{idg} is an alternative algorithm to convolutional gridding. In fact, it effects the same outcome---by convolving and resampling visibilities onto a regularised grid---but it does so by way of a detour through the image domain. It is on this detour that \textsc{idg} allows us to cheaply and efficiently apply $A$-term corrections, and which makes the algorithm a promising alternative to more expensive alternatives like AW-projection.

In this section we describe the \textsc{idg} algorithm following the description as set out in \citet{VanDerTol2018}, with additional notes to clarify and expand upon aspects of the algorithm.

\subsection{Convolutional gridding}

Our starting point in this derivation begins with convolutional gridding with AW-projection. 
In broad strokes, this algorithm works by: projecting visibilities throughout the $(u, v, w)$ volume onto the $w=0$ plane; correcting them for $A$-term effects; convolving this plane with an appropriate kernel; and finally resampling the visibilities as a regularised grid. From here, a dirty image can be constructed by an efficient \textsc{fft} into the image domain.

First, consider the interferometric measurement equation in discrete form as a sum over point sources (pixels, model components, etc.):

\begin{gather}
    V_j = \sum_i A_j(l, m) I(l, m) e^{-2 \pi i [u l_i+ v m_i + w n'_i]} \\
    \text{where} \quad n' = \sqrt{1 - l^2 - m^2} - 1 \nonumber
\end{gather}

These terms have their usual definition as defined, for example, in \citet{Thompson2017}. The $A$-term is parameterised by both $l$ and $m$ to indicate direction-dependence. Additionally, it has the subscript $j$ to indicate dependence on the $j$th visibility including its time of observation and baseline configuration.

Now we define $\tilde{V}(u, v) = \mathcal{F}\left\{T(l, m) \cdot I(l, m)\right\}$, that is, as the $A$-term-corrected visibilities along the plane $w=0$, and convolved with some taper $T$ that has desirable anti-aliasing properties. Following an identical procedure as in \citet{Cornwell2008}, we can express $\tilde{V}$ in terms of a convolution with the discrete $V_j$ samples and a series of kernels:

\begin{align}
    \tilde{V}(u, v) = &\sum_j V_j \delta(u - u_j, v - v_j) \nonumber \\
    & \ast \mathcal{F}\left\{T(l, m)\ A_j^{-1}(l, m)\ e^{2 \pi i w_j (n - 1)}\right\} \nonumber \\
    = &\sum_j V_j \delta(u - u_j, v - v_j) \ast C_j(u, v)
    \label{eqn:awprojection}
\end{align}

Where $C_j(u, v)$ is the Fourier transform of $c_j(l, m) = T(l, m) \cdot A_j^{-1}(l, m) \cdot e^{2 \pi i w_j (n - 1)}$. Each of the kernels has a different operation: the complex exponential projects the visibility onto the $w=0$ plane; the inverse $A$-term corrects for distortions to the visibility; and $T$ applies the anti-aliasing taper. In this step we also include a formalism that we will use later in the derivation of \textsc{idg}: we additionally multiply the discrete samples $V_j$ by the Dirac delta function centred at $(u_j, v_j)$, transforming the single sample point into a function over $(u, v)$.

AW-projection implements this equation and performs the (de)gridding operations by a series of convolution operations in visibility space. \textsc{idg}, on the other hand, uses \autoref{eqn:awprojection} as a starting point.

\subsection{Partitioning}

The first requirement of \textsc{idg} is that we partition the sampled visibilities into so-called `subgrids'.

A subgrid is a small subsection of $(u,v)$ space, centred at some arbitrary coordinate $(u_0, v_0, w_0)$, and spanning $L \times L$ pixels along the $u$ and $v$ dimensions, respectively. The subgrid inherits the same regularised grid spacing as the full visibility grid, and its centre snaps to one of these pixels. $L$ here is small, typically $\sim 100$ pixels or so, and so spans just a small section of the full visibility domain. For the meantime, we ignore the $w$ coordinate and set the subgrid origin to lie along $w = 0$.

We then proceed to partition the set of visibilities by assigning each visibility to exactly one subgrid. A visibility may be assigned to a subgrid if its $(u, v)$ coordinate lies within the span of the subgrid. Multiple visibilities may be assigned to the same subgrid. For any visibilities that do not lie within the span of a subgrid, it will be necessary to create one.

Importantly, this is a partitioning of visibilities, not of $(u, v)$ space. Depending on the partitioning algorithm used, it is possible that subgrids may overlap, but each visibility belongs to just one subgrid. By the same token, large swathes of $(u, v)$ space may be empty of subgrids if there are no samples made in those regions.

Having partitioned the full set of visibilities into subgrids, we can modify \autoref{eqn:awprojection} to be the sum of convolved subgrids. Consider for the moment only the $p$th subgrid. We can produce $\tilde{V}_p$ as the convolution of its member visibilities like so:

\begin{equation}
    \tilde{V_p}(u, v) = \sum_k V_k \delta(u - u_k, v - v_k) \ast C_k(u, v)
\end{equation}

where the subscript $k$ sums over only those visibilities belonging to the $p$th partition. To support the convolution kernel, subgrids must provide additional padding around their border.

Then, since convolution is distributive, the full set of gridded visibilities, $\tilde{V}(u, v)$, is simply the sum over all subgrids:

\begin{align}
    \tilde{V}(u, v) &= \sum_p \tilde{V_p}(u, v) \nonumber \\
    &= \sum_p \sum_k V_k \delta(u - u_{k}, v - v_{k}) \ast C_k(u, v)
\end{align}

\subsection{The origin shift}

The central `trick' of \textsc{idg} is the origin shift. This step drastically reduces the resolution of the visibility and image domains, and allows \textit{direct} imaging to be computationally feasible.

Let's begin by defining $\hat{V}_p$ as a shift of $\tilde{V}_p$ from its central coordinate $(u_{0p}, v_{0p})$ to the origin, which we can express using a Dirac delta function:\footnote{Using the property of Dirac delta functions where we can effect a coordinate translation by the application of a convolution: $f(x + a) = f(x) \ast \delta(x + a)$.}

\begin{align}
    \hat{V}_p(u, v) &= \tilde{V}_p(u + u_{0p}, v + v_{0p}) \nonumber \\
    &= \delta(u + u_{0p}, v + v_{0p}) \nonumber \\
    &\quad \ast \sum_k V_k \delta(u - u_{k}, v - v_{k})  \ast C_k(u, v) \nonumber \\
    &= \sum_k V_k \delta(u + u_{0p} - u_{k}, v + v_{0p} - v_{k}) \ast C_k(u, v)
    \label{eqn:Vhat}
\end{align}

We can then recover $\tilde{V}(u, v)$ simply by shifting the convolved subgrid back into place:

\begin{align}
    \tilde{V}(u, v) &= \sum_p \delta(u - u_{0p}, v - v_{0p}) \ast \hat{V}_p(u, v)
    \label{eqn:shiftback}
\end{align}

It's not obvious why one would go about the hassle of shifting the subgrid, only to shift it back again. However, let us consider $\hat{V}_p(u, v)$ by itself and define $\hat{I}_p(l, m)$ as the sky image of the shifted subgrid. These are related by discrete inverse Fourier transform:

\begin{align}
    \hat{I}_p(l, m) &= \mathcal{F}^{-1}\left\{ \hat{V}_p(u, v) \right\} \nonumber \\
    &= \sum_k V_k e^{2 \pi i \left[(u_k - u_{0p}) l + (v_k - v_{0p}) m\right]} \cdot c_k(l, m)
    \label{eqn:Ihat}
\end{align}

Here we've substituted in the expression for $\hat{V}_p$ given in \autoref{eqn:Vhat} and replaced the convolution with the multiplication of their Fourier counterparts. We have also applied the well-known result for the Fourier transform of the Dirac delta function.

The key insight is that now, because the $L_p \times L_p$ subgrid is small, then each of $(u_k - u_{0p})$ and $(v_k - v_{0p})$ are also small. That is, each visibility in the expression for $\hat{I}_p$ now behaves as if it were a very short baseline, mapping the sky at a correspondingly low resolution. As a result, it becomes possible to sample $\hat{I}_p$ with a small number of pixels and, in fact, we can critically sample $\hat{I}_p$ using a grid with a resolution of exactly $L_p \times L_p$. Therefore, provided $L_p$ is small, it now becomes computationally tractable that we can produce $\hat{I}_p$ by direct sum.

If we are able to cheaply compute $\hat{I}_p$ in this way, we can unwind the process and combine equations \ref{eqn:shiftback} and \ref{eqn:Ihat} to recover $\tilde{V}(u, v)$ as:

\begin{align}
    \tilde{V}(u, v) &= \sum_p \delta(u - u_{0p}, v - v_{0p}) \ast \mathcal{F} \left\{ \hat{I}_p(l, m) \right\}
\end{align}

Or in words: we compute $\hat{I}_p$ by direct sum, Fourier transform the grid back to produce $\hat{V}_p$, and then sum the subgrids onto $\tilde{V}$, having first shifted each back to its original position. This, more or less, is the \textsc{idg} algorithm.

\subsection{The \textit{w} coordinate}

There's one additional complication. Recall that we defined $c_k(l, m) = T(l, m) \cdot A_k^{-1}(l, m) \cdot e^{2 \pi i w_k (n - 1)}$, as a combination of an arbitrary taper, the $A$-term, and a $w$-projection term. This latter term allowed us to project a visibility sampled at some arbitrary $w_k$ onto the $w = 0$ plane. More generally, to project a visibility point onto an arbitrary $w = w_0$ plane, we would write this as:

\begin{equation}
    c_k(l, m) = T(l, m) \cdot A_k^{-1}(l, m) \cdot e^{2 \pi i (w_k - w_0) (n - 1)}
\end{equation}

If we include this expression explicitly in the equation for $\hat{I}_p(l, m)$, we find:

\begin{gather}
    \hat{I}_p(l, m) = \sum_k V_k e^{2 \pi i \phi} \cdot T(l, m) \cdot A_k^{-1}(l, m)  \label{eqn:withw} \\
    \text{where} \quad \phi = (u_k - u_{0p}) l + (v_k - v_{0p}) m \nonumber \\
    \qquad + (w_k - w_0)(n - 1) \nonumber
\end{gather}

The breakthrough in the \textsc{idg} algorithm is in being able to sample $\hat{I}_p(l, m)$ using the low-resolution $L_p \times L_p$ grid. However, in this explicit form, it becomes apparent that to ensure that the function remains critically sampled at this resolution it is necessary to place a limitation on the size of $|w_k - w_0| < w_\text{max}$.\footnote{Calculating this limit is somewhat tricky, since $n$ is a (non-linear) function of $l, m$. Pigi uses an heuristic that calculates the smallest possible fringe pattern produced by this $w$ term given any of the valid values of $u$, $v$ and $n$ over the field. We can then determine a maximum allowable limit $|w_k - w_0| < w_\text{max}$ that ensures this fringe is critically sampled everywhere.}

This value for $w_\text{max}$ affects partitioning. In the partitioning step we can no longer set $w_0 = 0$ for all subgrids. Instead, we must assign a particular visibility to a subgrid having a central coordinate $(u_0, v_0, w_0)$ based on two criteria: as before, the visibility's $(u_k, v_k)$ coordinate must lie within the span of the $L \times L$ subgrid, but additionally the visibility must now be `close by' in the $w$ direction such that $|w_k - w_0| < w_\text{max}$.

This solution leads quite naturally a hybrid algorithm that combines \textsc{idg} with $w$-stacking \citep[see e.g.][]{Offringa2014}. In Pigi, the visibilities are partitioned by first discretising $w$ space into layers $\{ w_\text{max}, 3 w_\text{max}, 5 w_\text{max}, ...\}$.\footnote{Note that we can reduce the number of $w$-layers by forcing visibilities to have a positive $w$ term. This ability can be derived from the measurement equation, where it follows that $V(u, v, w) = V(-u, -v, -w)^\dagger$.} All subgrids snap exactly to one of these layers. The \textsc{idg} algorithm produces visibilities gridded onto one of these $w$ layers, $\tilde{V}_w(u, v)$, and as with the $w$-stacking algorithm, the final step is to Fourier transform each layer into the image space, to apply a correction for the offset from $w = 0$, and then sum. That is,

\begin{gather}
    I(l, m) = \nicefrac{1}{T(l, m)}\sum_w I_w(l, m) \cdot e^{2 \pi i w (n - 1)} \label{eqn:wstack} \\
    \text{where} \quad I_w(l, m) = \mathcal{F}\left\{ \tilde{V}_w(u, v) \right\} \nonumber
\end{gather}

\subsection{The algorithm}
\label{sec:algorithm}

Putting this altogether, we now demonstrate the \textsc{idg} algorithm in the imaging direction:

\begin{enumerate}
    \item Partition the set of visibilities by assigning each to an $L \times L$ pixel subgrid spanning some small range in $(u, v)$, snapped to some nearby $w$ layer, and indexed by $p$.
    \item For the $p$th subgrid:
    \begin{enumerate}
        \item \textbf{Gridder kernel:} Compute each value of $\hat{I}_p$ by directly summing each of the assigned visibilities and multiplying through by the taper $T(l, m)$, as per \autoref{eqn:withw}.
        \item Fourier transform $\hat{I}_p$ to produce $\hat{V}_p$.
        \item \textbf{Adder kernel:} Unshift the subgrid back to its central coordinate and sum with other subgrids at the same $w$-layer, producing $\tilde{V}_w(u, v)$, as per \autoref{eqn:shiftback}.
    \end{enumerate}
    \item The visibilities are now fully gridded onto multiple $w$-layers. Then for each $w$-layer we perform $w$-stacking, as per \autoref{eqn:wstack}.
\end{enumerate}

In the opposite direction, we can predict by reversing the order of each of the gridding steps and inversing the operation:

\begin{enumerate}
    \item Partition the set of visibilities by assigning each to an $L \times L$ pixel subgrid spanning some small range in $(u, v)$, snapped to some nearby $w$ layer, and indexed by $p$.
    \item Apply the taper $T(l, m)$ to the image $I(l, m)$ and then perform inverse $w$-stacking. Namely, for each $w$-layer, find $\tilde{V}_w(u, v)$ as:
    \begin{gather}
        \tilde{V}_w(u, v) = \mathcal{F}\left\{  I(l, m) \cdot T(l, m) \cdot e^{-2 \pi i w (n - 1)} \right \}
        \label{eqn:wunstack}
    \end{gather}
    
    \item For the $p$th subgrid:
    \begin{enumerate}
        \item \textbf{Splitter kernel:} Copy the the overlapping visibility values from $\tilde{V}_w$ to $\hat{V}_p$. 
        \item Inverse Fourier transform $\hat{V}_p$ to produce $\hat{I}_p$.
        \item \textbf{Degridder kernel:} For each visibility assigned to the subgrid, indexed by $k$, compute the sum over each pixel, indexed by $i$:
        \begin{equation}
            V_k = \sum_i \frac{ \hat{I}_p(l_i, m_i) A_k(l_i, m_i)}{ T(l_i, m_i) } e^{-2 \pi i \phi}
            \label{eqn:degridder}
        \end{equation}
    \end{enumerate}
\end{enumerate}

\section{Implementation}

Pigi is written in C++20 using the Heterogeneous-computing Interface for Portability (\textsc{hip}) \textsc{api} to interface with \textsc{gpu} devices from both \textsc{nvidia} and \textsc{amd}. The focus in developing Pigi has been on the inversion and prediction routines and their respective kernels, but Pigi also has a simple Cotton-Schwab \citep{Schwab1984} multi-frequency cleaning implementation (following \citet{Offringa2017}) and the ability to perform full deconvolution. The codebase is reasonably compact, with the inversion and prediction routines and their kernels amounting to about 1500 lines of heavily templated code, and the full codebase at about 7000 lines.

In this section we expand upon the concrete implementation of the \textsc{idg} algorithm in Pigi.

\subsection{Partitioning}

\begin{figure*}
    \centering
    \includegraphics[width=0.7\linewidth,clip,trim={2cm 0 2cm 0}]{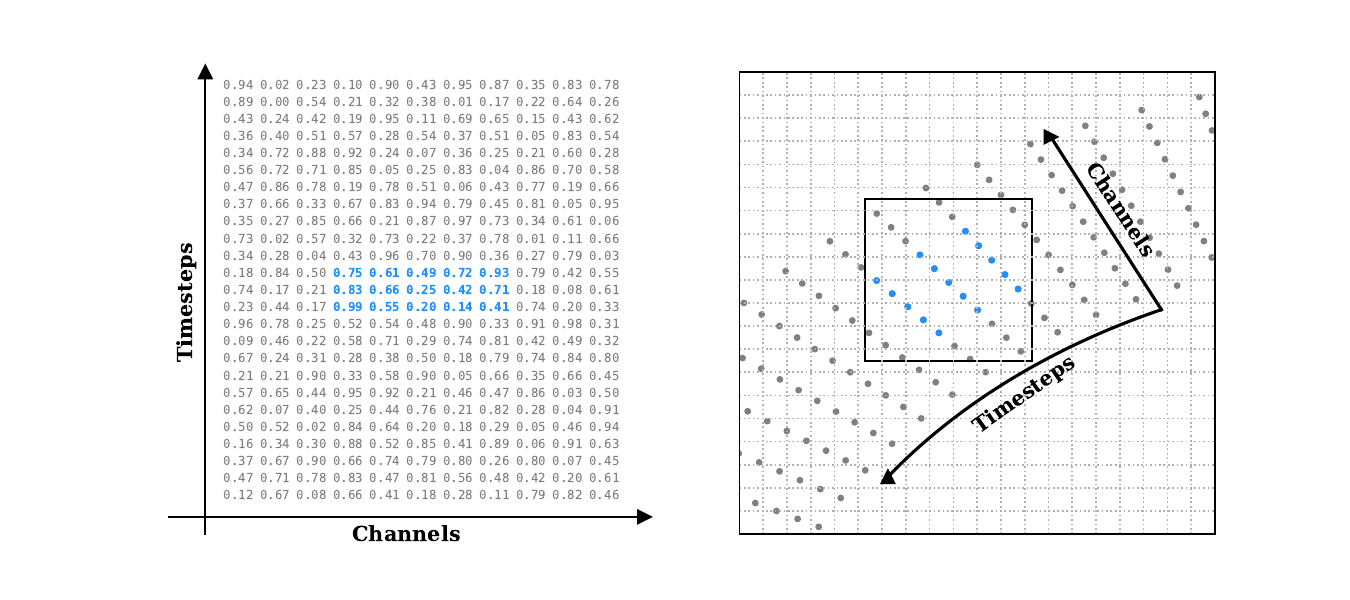}
    \caption{A representation of the partitioning algorithm used in Pigi. The diagram on the left represents the visibility data for a particular baseline pair, arranged in a two-dimensional table spanning frequency channels and timesteps. The diagram on the right shows the location of those same visibilities on the $uv$ grid. Partitioning data on the $uv$ grid corresponds to decomposing the visibility table into small subtables.}
    \label{fig:partition}
\end{figure*}

In both gridding and degridding, the first step is to partition the data set into subgrids.

For a single baseline pair, typical interferometric data can be arranged as a table having rows that correspond to timesteps and columns that correspond to frequency channelisation. As we move across channels, the $(u,v,w)$ coordinates of a visibility stretch as a function of the observation frequency, and as we move down a row, they rotate slowly in space at the speed of the Earth's rotation.

In Pigi, we use this correspondence to aid in partitioning, whereby partitioning in $uv$ space amounts to partitioning the visibility data into a series of subtables spanning some small time and channel range (see \autoref{fig:partition}). More specifically, Pigi partitions the full table in a single pass by: partitioning a single timestep into a series of channel ranges that each `comfortably' fit into a subgrid, and then proceeding to expand the subtable with additional timesteps until the span of the subgrid is exceeded. `Comfortably' in this sense entails leaving some room to grow, and avoids an excess of subgrids that are composed of a single timestep. 

This algorithm stands in contrast to an earlier `greedy' algorithm that Pigi originally implemented. In this earlier algorithm, we only created a new subgrid if the current visibility could not be assigned to any of the existing candidate subgrids. While this had the advantage of ensuring the number of subgrids was kept at a minimum, the subtable approach brings two important advantages.

Firstly, subtable partitioning is more efficient. Assuming channels are ordered by frequency, only the `corners' of a subtable must be checked to ensure it remains sized within the subgrid bounds which dramatically reduces the number number of tests required. In contrast, the greedy method requires a full subgrid search for every visibility.

Secondly, subtable partitioning keeps data `nearby' each other in memory and makes sending visibility data to the \textsc{gpu} efficient. Greedy partitioning, on the other hand, results in high memory fragmentation and incurs a significant associated performance penalty.

\subsection{Kernels}

In \autoref{sec:algorithm}, we've identified four named kernels required for \textsc{idg}. There are additionally a number of other small ad-hoc kernels used for things like \textsc{fft} normalisation or applying the $\Delta l, \Delta m$ shift used in multi-field imaging. Of all these kernels, however, it is the gridder and degridder kernels that are most computationally expensive and dominate the runtime. In this section, we provide some implementation details on each of the gridder and degridder kernels.

\subsubsection{Gridder}

The gridder must perform a direct Fourier transform for each subgrid, as per equation \autoref{eqn:Ihat}. The complexity of this algorithm is $\mathcal{O}(N L^2)$, where $N$ is the number of visibilities to be gridded and $L$ is the span of one dimension of the subgrid.

Pigi's gridder implementation processes the data in batches, sized so as to fit in the \textsc{gpu} memory. Each batch consists of a set of subgrids (a contiguous stack of $L \times L$ arrays), associated metadata, and a contiguous set of visibilities arranged as a two-dimensional grid spanned by timesteps and channels.

The gridder is distributed over the \textsc{gpu} along two dimensions. The outer y-dimension pertains to the subgrid index, allocating one block per subgrid and with a thread size of 1. The inner x-dimension is the principal axis of parallelisation and pertains to each of the subgrid pixels: each block has a fixed number of threads,\footnote{This is a tunable parameter per device. For example, on the \textsc{nvidia} \textsc{a100} a thread size of 128 results in best performance. On the \textsc{amd} \textsc{mi250x}, on the other hand, 64 threads is optimal.} and there are as many blocks as required to process each subgrid pixel.

In this way, each thread is responsible for a single pixel belonging to one of the subgrids, and performs its own independent reduction over the visibilities.

Visibility data is retrieved from global memory and inserted into a shared memory buffer using coalesced reads across the block. This staging of data is currently handled manually as \textsc{hip} does not yet support \textsc{cuda}'s newer \texttt{cooperative\_groups}. During this data loading stage, some initial data processing, such as weighting, is also performed by each thread prior to writing to shared memory. Each thread then sums over the current set of visibilities in the shared memory, iteratively refilling the shared memory buffer with the next set of visibilities, and continuing until they are finally exhausted.

In addition to staging data into shared memory, some other optimisations of note are:

\begin{enumerate}
    \item The inner loop involves the computation of the sine and cosine of the phase term. In single precision mode, we use the \texttt{\_\_sincosf()} function, which computes both values at once with an increased error tolerance compared to the more accurate \texttt{sincosf()} function. This function leads to a significant improvement in kernel throughput. The trade-off is a reduced precision of the gridder, which leads to an approximately twofold increase in the error.
    \item Fused multiply add (fma) instructions combine a multiplication and addition into a single instruction. The compiler is normally clever enough to insert these, however in the case of complex numbers we found it necessary to do this explicitly by writing our own fused multiply accumulate function.
    \item Thread coarsening along the \textsc{gpu}'s x-dimension by a factor of 4 (i.e.\@ processing 4 subgrid pixels per thread) marginally improves the computational intensity of the kernel whilst still maintaining good \textsc{gpu} occupancy and despite increased register pressure.
    \item By processing visibilities by row (i.e.\@ all having the same baseline) we are able to pre-compute the phase term for all channels, having dimensionality meters. The inner loop then need only divide this term by the wavelength of the channel before computing the complex phasor, reducing the number of inner loop instructions. This optimization, however, will perform poorly in the case of narrow band imaging.
\end{enumerate} 

\subsubsection{Degridder}

The degridder performs the inverse operation to the gridder, as per equation \autoref{eqn:degridder}. As with the gridder, its complexity is $\mathcal{O}(N L^2)$.

The degridder performs a reduction over each subgrid pixel onto each of its associated visibilities. The most natural implementation of the kernel would be to mirror this structure: to assign each \textsc{gpu} thread to a single visibility and let it perform an independent reduction over each of its associated subgrid pixels. However, unlike the subgrid size, which is constant and known ahead of time, the number of visibilities assigned to each subgrid is variable. Moreover, in some imaging configurations, a substantial fraction of subgrids may have a visibility occupancy of only a few tens of visibilities. These variable requirements don't map well onto a statically sized \textsc{gpu} grid and result in an uneven tail of work that can increasingly dominate the runtime in those cases where subgrid occupancy becomes low.

Pigi, instead, distributes the degridder over the \textsc{gpu} identically to the gridder: the outer y-dimension allocates one block per subgrid, and the inner x-dimension allocates blocks each having 128 threads to each of the subgrid pixels. This, however, means that the reduction can no longer be performed independently by each thread and instead, threads cooperate at the warp level to perform a reduction over each of their assigned pixels.\footnote{Warp-level reduction is implemented by constructing a registry cache amongst warp members and then daisy-chaining visibilities within the warp. When the full cycle is complete each thread finally performs an atomic add operation back to global memory. The reduction is performed at the warp level using a registry cache, as opposed to a shared memory cache, to avoid the more costly atomic operations that would be required in the latter instance.} In cases with high subgrid occupancy, this overhead of warp-level reduction results in a kernel that is about 20\% slower than the more `natural' implementation; however, in those cases where subgrid occupancy is less than a few hundred visibilities, this implementation avoids runaway performance degradation.\footnote{A future optimisation candidate would be to swap out each kernel design based on a subgrid occupancy heuristic.}

As with the gridder, the degridder takes advantage of each of the same aforementioned optimisations. However, the warp-level reduction requires processing rows having a channel count that is an even multiple of the warp size, and any tail will result in wasted instructions. This fact, combined with the overhead of warp-level synchronisation primitives, leads to overall lower throughput than the gridder (see \autoref{sec:kernelthroughput}).

\subsection{\textit{A}-term application}

In equation \ref{eqn:withw}, we've indexed $A$ by $k$ to stress the dependency of the $A$-term on the $k$th visibility and the possibility of the baseline- or time- dependence of this term. This potential variability of the $A$-term forces it \textit{inside} the summation and adds significant computational and memory overhead, especially when the sum is considered in its full polarised form where, using the Jones formalism, each of $A$ and $V$ are complex-valued $2 \times 2$ matrices.

The mitigation is simple, however: if the $A$-term is constant across the sum, then its application can be deferred until afterwards. Pigi thus includes additional constraints during the partitioning phase, whereby visibilities can be associated with a subgrid only if they also all share the same $A$-terms.

One other implementation detail of note is that, to minimise noise, the power of the $A$-term must be normalised at this point and later corrected in the combined image. This corrects for beam nulls, for example, whose noise would otherwise be unduly amplified. Pigi mitigates this by normalising $\hat{I}$ by the Frobenius norm of the product $A_1^{-1} \left( A_2^{-1} \right)^{\dagger}$. When inversion is complete, the final image must then be corrected by the weighted average of this normalisation factor across all subgrids.

\subsection{\textit{w}-stacking and data batching}
\label{sec:wstacking}

Pigi sends data to the \textsc{gpu} in batches that are sized to ensure memory usage remains within limits. To ensure efficient memory transfers between host and \textsc{gpu}, a batch consists of a contiguous region of visibility memory. The memory usage of a batch is dominated by two objects: the subgrid stack, and their associated visibilities. However, the memory allocated to a batch competes with the memory allocation of the master grid and its \textsc{fft} plan. As the image size becomes larger and the master grid similarly increases in the size, batches will become smaller and more numerous.

For each batch, in addition to running the (de)gridder kernels, Pigi runs the $w$-stacking routine, which is implemented naively as per equations \ref{eqn:wstack} and \ref{eqn:wunstack}. Part of this process requires shifting the master grid between $w$-layers. This process has a computational cost that is dominated by a pair of \textsc{fft} operations, that is $\mathcal{O}(N \log{N})$, where $N$ is the number of image pixels.

This $w$-stacking routine represents a fixed-cost for each batch. Since the contents of a batch is determined only by the requirement that the visibilities are contiguous in memory, it is likely that each batch is composed of subgrids assigned to multiple, and typically all $w$-layers. Performing $w$-stacking on each batch therefore requires iterating through each of these $w$-layers, and has a fixed cost that depends only on the image size. This has important effects on performance, especially when memory constraints force batch sizes to be small, which we discuss in \autoref{sec:subgridcount}.

\subsection{Multiprocessing model and\\
data ownership}
\label{sec:mpi}

Pigi uses \textsc{mpi} to distribute data and, in turn, computation, over multiple nodes and \textsc{gpu}s. For simplicity of implementation, Pigi is always required to run under \textsc{mpi} and requires $(N + 1)$ ranks, with $N$ workers for each data subset and an additional rank required for the queen process. The \textsc{mpi} implementation is free to locate workers as it pleases, allowing for both single and multi-node operations.

Pigi assigns each worker a unique subset of data which it owns for the life of the program. Each worker initialises itself by loading its assigned data into memory, and then the queen calls upon the workers to invert or predict the data as needed. The benefit of this model, besides its simplicity, is that it minimises disk read/write operations in those cases where it is possible to load data fully into memory. In scenarios with large supercomputing resources, distributing a data set over enough nodes so as to allow in-memory operation leads to a significant performance increase. In cases where this is not possible, Pigi memory maps data to a temporary file.

Presently, data is partitioned across ranks using a \texttt{channels-out} parameter, whereby the full channel range is decomposed into $N$ subranges. The \texttt{channels-out} configuration parameter arises separately to address spectral changes in wideband deconvolution \citep{Offringa2017} but is repurposed here also a means to distribute data. In the future, we envision a similar \texttt{times-out} parameter to provide an identical data partitioning axis in the time domain.

\textsc{gpu}s are allocated in a round robin fashion to ranks, and each rank will use just the one \textsc{gpu} over the lifetime of inversion and prediction requests from the queen. This is a coarse allocation of \textsc{gpu} resources and may be suboptimal in cases where work is unbalanced among ranks, however it is currently preferred given its simplicity of implementation. 

The queen is responsible for coordinating work amongst workers, collecting and merging the results, as well as deconvolution. Deconvolution is presently implemented as a Cotton-Schwab clean (with some additional optimisations to minimise the search space) and is fundamentally serial in its algorithm. Typically the final major cycle of cleaning can take a significant amount of time, and this is time that the workers are left idle despite occupying \textsc{gpu} and memory resources. Future work could look at improving the speed of cleaning, such as aggressive masking in the final rounds of cleaning, or pursuing deconvolution algorithms that can better make use of multi-node resources.

See \autoref{sec:scalability} for performance benchmarks of Pigi in multinode environments. 

\section{Accuracy}

\begin{figure*}
    \centering
    \includegraphics[width=1\linewidth]{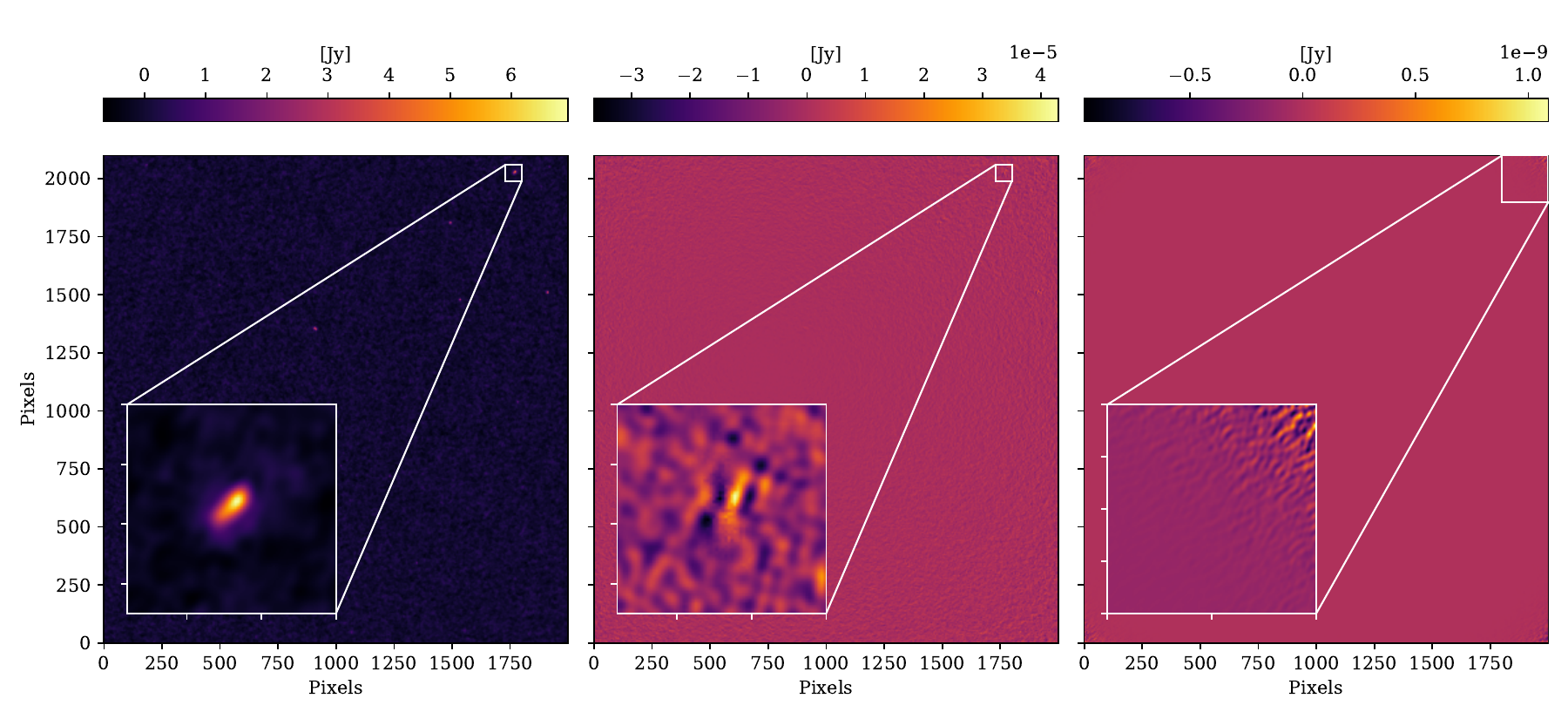}
    \caption{A comparison of errors arising from single (32 bit) and double (64 bit) floating point precision with spatially uniform $A$-terms. \textit{Left:} The original dirty image. \textit{Middle:} the error produced using single precision with peak error \SI{43}{\micro \jansky} and \textsc{rms} error \SI{2.5}{\micro \jansky}. \textit{Right:} the error produced using double precision with peak error \SI{1.1}{\nano \jansky} and \textsc{rms} error \SI{15}{\pico \jansky}. Note that the types of error differ: the dominant sources of error in 32 bit mode are phase errors around the peaks of bright sources, as a result of the reduced accuracy of trigonometric functions; whilst in 64 bit mode, the dominant error is related to the taper and is largest at the extreme corners of the image.}
    \label{fig:accuracy}
\end{figure*}

The \textsc{idg} algorithm achieves high levels of accuracy. In this section we discuss the types of errors we have observed with Pigi and the effect of some of the tunable configuration parameters. Throughout, we will be comparing the results of imaging a calibrated 2-minute \textsc{mwa} snapshot produced by direct Fourier transform at double precision---the `true' image---with those created by Pigi.\footnote{Unless otherwise stated, the Pigi imaging configuration uses a pixel scale of \SI{15}{\arcsecond}, image padding factor of 1.5, subgrid size of 96 pixels, a kernel padding of 18 pixels, and a spatially uniform $A$-term set to the identity matrix.}

\subsection{Floating point precision}

We start by discussing the choice of floating point precision, namely between single (32 bit) and double (64 bit) precision modes. Pigi parameterises the choice of floating point precision throughout its codebase which allows for modifying the precision whilst ensuring full code reuse. In \autoref{fig:accuracy}, we present a set of 2k $\times$ 2k pixel images showing the true image on the left, and in the middle and right we show the difference images produced by Pigi at single and double precision, respectively. The true image has a peak value of \SI{6.9}{\jansky} and root mean squared (\textsc{rms}) value across the map of \SI{170}{\milli \jansky}. The Pigi difference images have \textsc{rms} errors of \SI{2.5}{\micro \jansky} and \SI{15}{\pico \jansky}, for each of the single and double precision imaging modes, giving signal to noise (\textsc{snr}) values of \SI{48.4}{\decibel} and \SI{100.5}{\decibel}, respectively.

The source of the errors differ for each of the precision modes. At single precision, the dominant source of error arises from inaccuracies in the trigonometric function used to calculate the complex phase term, namely \texttt{\_\_sincosf()}. The result of this inaccuracy is an error that is distributed throughout the image, especially around bright sources. As an example, in the left inset we show a bright source in the true image, and in the middle inset we can see the residual in and around this source caused by the trigonometric imprecision. At double precision, on the other hand, the dominant source of error appears only at the corner edges of the image. In the right inset we show an example of this, where the magnitude of the error rises as a function of radial distance from the image centre. This error is due to numerical instability in the application and removal of the taper, which when using a padding factor of 1.5 has a value on the order of $10^{-8}$ at the corner edge.

\subsection{Taper and kernel padding}

The choice of taper function and the amount of kernel padding are also key interrelated factors that affect the accuracy of the final image. Pigi implements both the Kaiser Bessel (\textsc{kb}) and Prolate Spheroidal Wave Function (\textsc{pswf}). The \textsc{kb} function is simple to implement and has support in the C++ standard library. The more complex \textsc{pswf} can be shown to maximise the energy in the main lobe whilst requiring limited support in the visibility domain and is the mainstay taper function used in many radio astronomy imaging applications (see e.g.\@ \citet{VanDerTol2018} for a more complete discussion).

\begin{figure}
\centering
\includegraphics[width=1\linewidth]{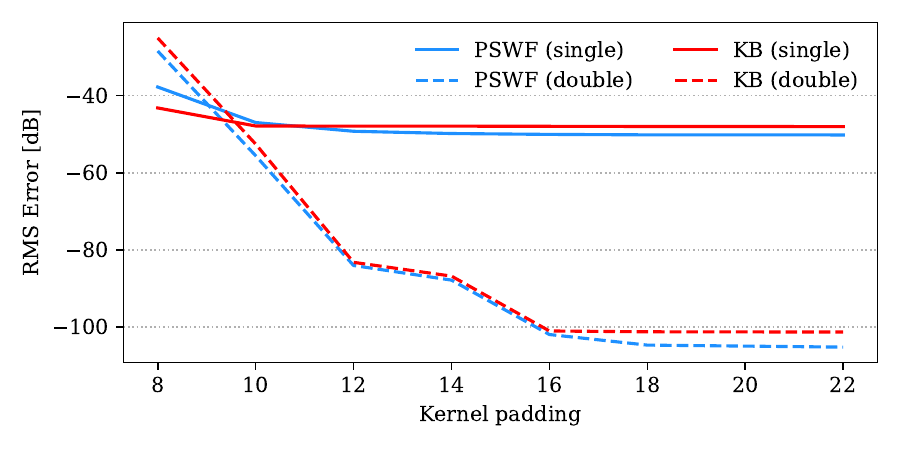}
\caption{A comparison of the \textsc{rms} error for varying kernel padding values and taper types when imaging an \textsc{mwa} snapshot observation. The error is measured in comparison to an image produced by direct Fourier transform.}
\label{fig:taper}
\end{figure}

In \autoref{fig:taper} we show the \textsc{rms} error across a 2k $\times$ 2k pixel image (using a 96 pixel subgrid) as a function of kernel padding, and for each of the implemented taper functions. Whilst the \textsc{pswf} can produce errors marginally lower than the \textsc{kb} window, in practice the choice of taper has very little effect on the achieved error. At single precision and for kernel padding values greater than 12 pixels, the error reaches a minimum plateau and is instead dominated by the trigonometric errors identified earlier. Likewise, at double precision the minimum plateau is reached with a padding size of 16 pixels, at which point the numerical instability arising from the taper edges dominates the error.

\subsection{\textit{A}-term interpolation errors}

The chosen subgrid size has an important effect on the accuracy of the $A$-term function. The $L \times L$ subgrid size determines the number of samples taken of the underlying $A$-term function; in between these sample points the \textsc{idg} algorithm performs the equivalent of sinc interpolation. To minimize inaccuracies, it is essential to sample the $A$-term function at a sufficient density to capture its spatial detail.

\begin{figure*}
    \centering
    \includegraphics[width=1\linewidth]{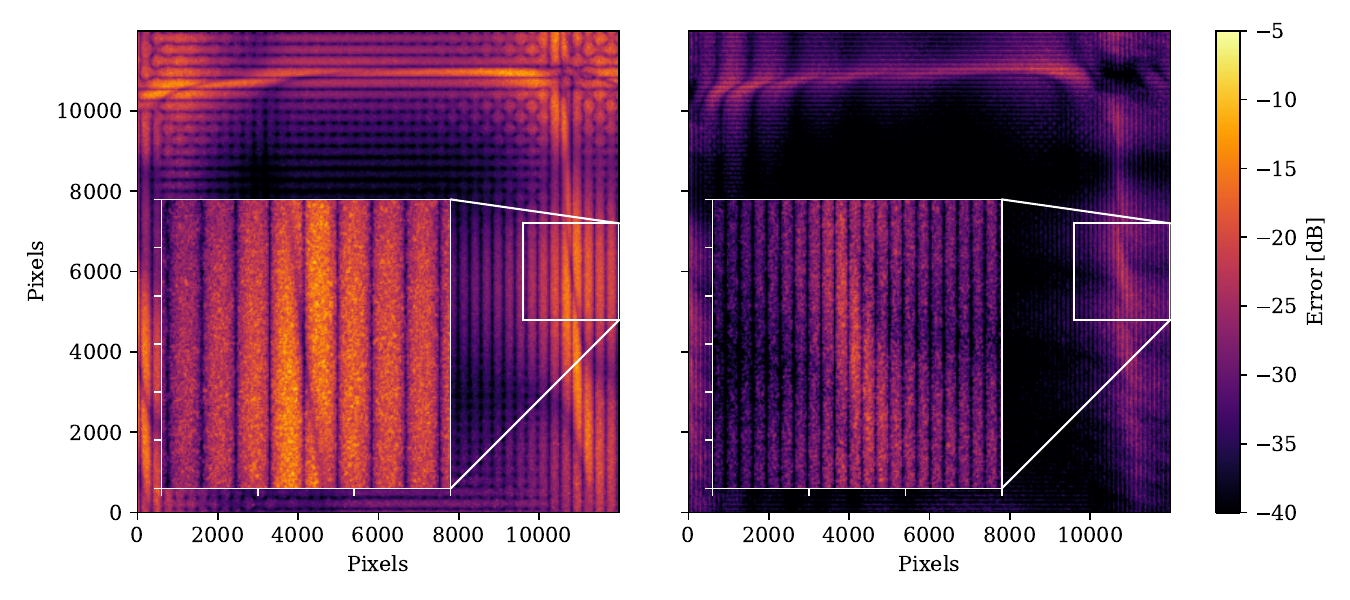}
    \caption{A widefield $12000 \times 12000$ pixel image showing the primary beam error in and around beam nulls using a subgrid of size 64 pixels (left) and 160 pixels (right). More frequent sampling of the beam narrows the fringe pattern and reduces the error.}.
    \label{fig:widefield}
\end{figure*}

Until now, this dicussion on accuracy has considered only the case of a spatially uniform $A$-term. Now instead consider, as an example, the \textsc{mwa} primary beam. This is a smooth, slowly-varying function over the majority of its primary lobe, however in and around its nulls the primary beam changes orders of magnitude quite rapidly. \autoref{fig:widefield} shows the imaging errors across a widefield 12k $\times$ 12k pixel image that includes the \textsc{mwa} beam nulls.\footnote{To make this comparison, both the true image and the Pigi images have been modified, prior to differencing, to have constant sensitivity across the field, i.e.\@ normalized by the beam power.} The images are produced using subgrid sizes of 64 pixels (left) and 160 pixels (right). In the central 4k $\times$ 4k pixel region, where the beam is smoothly varying, the \textsc{rms} error is \SI{-32}{\decibel} and \SI{-40}{\decibel} for the $L=64$ and $L=160$ pixel configurations respectively. This error is significantly higher, however, along the primary beam null regions, with peak errors of \SI{-4.6}{\decibel} and \SI{-10.9}{\decibel}. The inset regions show the characteristic ridge pattern of the error, with the error minimised at or near the sample points and peaking in the interpolated region. Increasing the subgrid size has the effect to reduce both the width of these ridges as well as magnitude of the error.

In the case of the \textsc{mwa} beam, these errors are acceptable as they are constrained within null regions and otherwise do not affect deconvolution. However there are classes of $A$-term functions, for example ionospheric corrections, that have small-scale structure throughout the image and the effectiveness of their correction will be constrained by the limits of sinc interpolation across the subgrid.

\section{Performance}

\subsection{Hardware setup}

In this section we will be measuring the performance of Pigi across two systems as shown in \autoref{table:hardware}, comparing the performance of hardware from both \textsc{amd} and \textsc{nvidia}.

\begin{table*}[]
    \centering
    \small{
        \begin{tabular}{llllll}
            \toprule
            \textsc{\textbf{gpu}} & \textbf{Mem} & \textbf{Peak$^a$} & \textbf{Bandwidth$^b$} & \textsc{\textbf{cpu}} & \textbf{Mem} \\
            & [\SI{}{\giga \byte}] & [\SI{}{\tera \flop \per \second}] & [\SI{}{\giga \byte \per \second}] & & [\SI{}{\giga \byte}] \\
            \midrule
            \textsc{nvidia} \textsc{a100} ($\times 4)$ & 40 & 15.07 & 1549 & \textsc{amd} EPYC ($\times64$) & 512 \\
            \textsc{amd} \textsc{mi250x} ($\times 4)$ & 64$^c$ & 21.74$^c$ & 1380$^c$ & \textsc{amd} Trento ($\times64$) & 256 \\
            \bottomrule \\
            \multicolumn{6}{l}{
                \footnotesize
                \begin{tabular}[b]{l}
                $^a$ Peak single precision performance assuming purely \textsc{fma} instructions with a weighting of 2 as empirically\\
                measured at base clock speed.\\
                $^b$ Global memory bandwidth. \\
                $^c$ Each \textsc{mi250x} exposes two logical cores; these values are measured per core.
                \end{tabular}
            }
        \end{tabular}
    }

    \caption{Hardware setup showing the \textsc{nvidia} and \textsc{amd} accelerators used in performance testing. Peak and bandwidth values are measured empirically.}
    \label{table:hardware}
\end{table*}

The first system is a single machine hosting four \textsc{nvidia} \textsc{a100} \textsc{sxm} \textsc{gpu}s, each with 40 GB of device memory and connected via a \textsc{pci}e4 interconnect. The second system is provided as part of Setonix supercomputer run by the Pawsey Supercomputing Centre. Each node in Setonix has 4 \textsc{amd} \textsc{mi250x} \textsc{gpu}s, each with 128 GB device memory and connected via \textsc{amd} InfinityFabric. The \textsc{mi250x} exposes two logical cores, giving the appearance of 8 \textsc{gpu}s per node, and all benchmarks here are made using just one of these logical cores.

Disk read/write operations vary markedly between the two systems, with the \textsc{a100} having local SSD storage and the \textsc{mi250x} using a Lustre network filesystem. In these performance benchmarks, the disk read/write operations are excluded from measurements. 

\subsection{Rooflines and kernel profiling}

\begin{figure*}
    \begin{subfigure}{0.49\textwidth}
        \centering
        \includegraphics[width=1\linewidth]{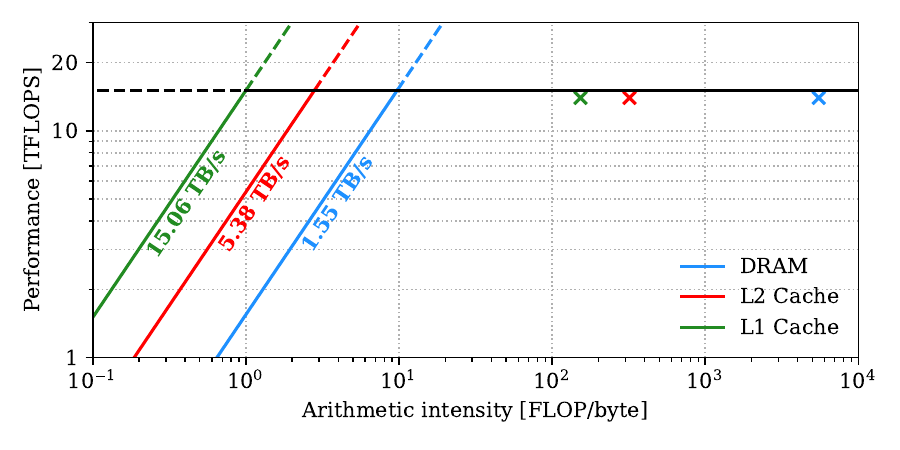}
        \caption{\textsc{nvidia} \textsc{a100}: \SI{13.96}{\tera \flop \per \second} (92.6\%)}
        \label{fig:roofline-gridder}
    \end{subfigure}
    \begin{subfigure}{0.49\textwidth}
        \centering
        \includegraphics[width=1\linewidth]{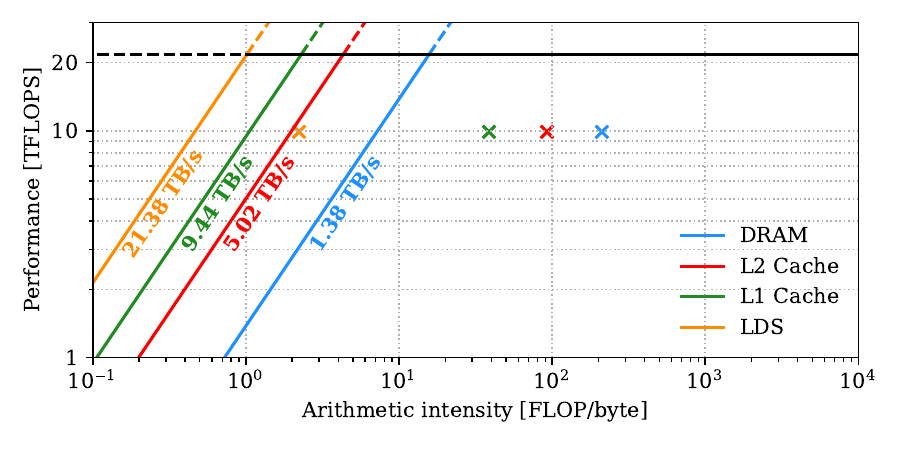}
        \caption{\textsc{amd} \textsc{mi250x}: \SI{9.90}{\tera \flop \per \second} (45.6\%)}
        \label{fig:roofline-mi250-gridder}
    \end{subfigure}
    \caption{Roofline plots for the gridder kernel (using single precision) on both hardware types. Despite the \textsc{amd} system having higher performance capacity, the kernel achieves poor floating point utilisation resulting in an overall lower compute performance.}
    \label{fig:roofline}
\end{figure*}

\begin{figure}
    \centering
    \includegraphics[width=0.8\linewidth,clip,trim={0.4cm 0 0.2cm 0}]{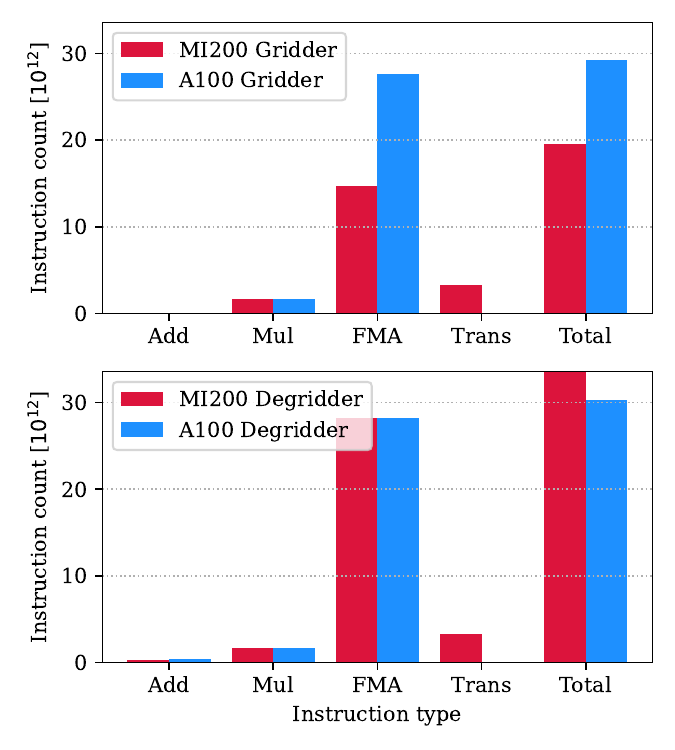}
    \caption{The floating instruction decomposition for the (de)gridder kernels on each of the \textsc{a100} and \textsc{mi250x}. All devices and kernels show strong utilisation of the more efficient \textsc{fma} instruction. Note that \textsc{trans} refers to the transcendental operations and is the \textsc{roc}m category for intrinsic functions such as trigonometric functions, square roots, etc.; such operations are handled by a separate pipeline on the \textsc{a100}.}
    \label{fig:instructions}
\end{figure}

Roofline plots are an important tool to discriminate between different types of hardware limitations, where each of the roofs indicate different types of a computational or memory bandwidth thresholds. \autoref{fig:roofline} shows roofline plots for the gridder kernels on both the \textsc{a100} and \textsc{mi250x} accelerators. The horizontal roofs correspond to peak floating point performance, whilst the slanted roofs correspond to different types of \textsc{gpu} memory bandwidths---global (\textsc{dram}), \textsc{l1} and \textsc{l2} caches---as well as shared memory bandwidth (\textsc{lds}) which we have only measured for the \textsc{mi250x}.

Despite being a more capable device than the \textsc{a100}, the \textsc{mi250x} suffers from poor floating point utilisation for both gridder and degridder kernels, and its performance suffers a result. On the \textsc{mi250x}, the gridder kernel achieves just 45.6\% utilisation, and for the degridder it achieves a similarly poor 56.8\% utilisation. The \textsc{a100}, on the other hand, achieves greater than 90\% utilisation for both kernels. For both devices, the kernels are situated in the computationally-bound region of the roofline plots, and neither suffers from significant memory bandwidth limitations.

The floating point instruction makeup could provide an indication as to why the \textsc{mi250x} achieves such poor utilisation, especially if the efficient fused multiply add (\textsc{fma}) instructions are poorly used. \autoref{fig:instructions} shows that for both kernels and on both architectures, however, we see that the more efficient \textsc{fma} instructions indeed dominate. The proportion of \textsc{fma} instructions is higher for the \textsc{a100} at $\sim92$\% compared to the \textsc{mi250x} at $\sim76$\%, owing principally to the fact that transcendental functions, such as trigonometry functions, are handled by a separate pipeline on the \textsc{a100}. This difference, however, doesn't fully account for the poor performance of the \textsc{mi250x}.

\subsection{Kernel throughput}
\label{sec:kernelthroughput}

\begin{figure*}
    \begin{subfigure}{0.49\textwidth}
        \centering
        \includegraphics[width=1\linewidth]{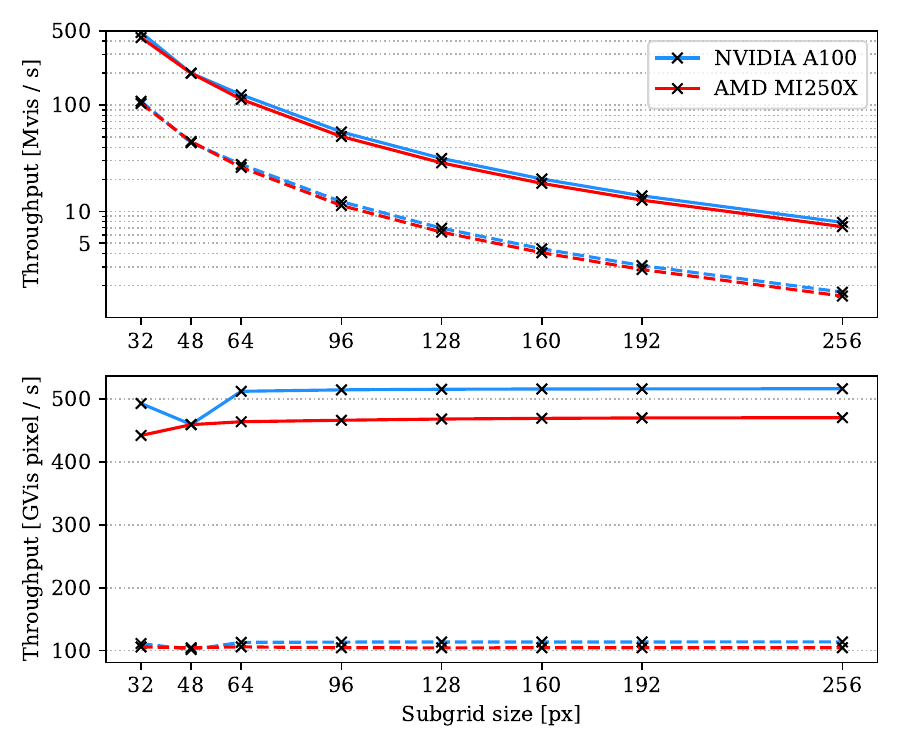}
        \caption{Gridder}
    \end{subfigure}
    \begin{subfigure}{0.49\textwidth}
        \centering
        \includegraphics[width=1\linewidth]{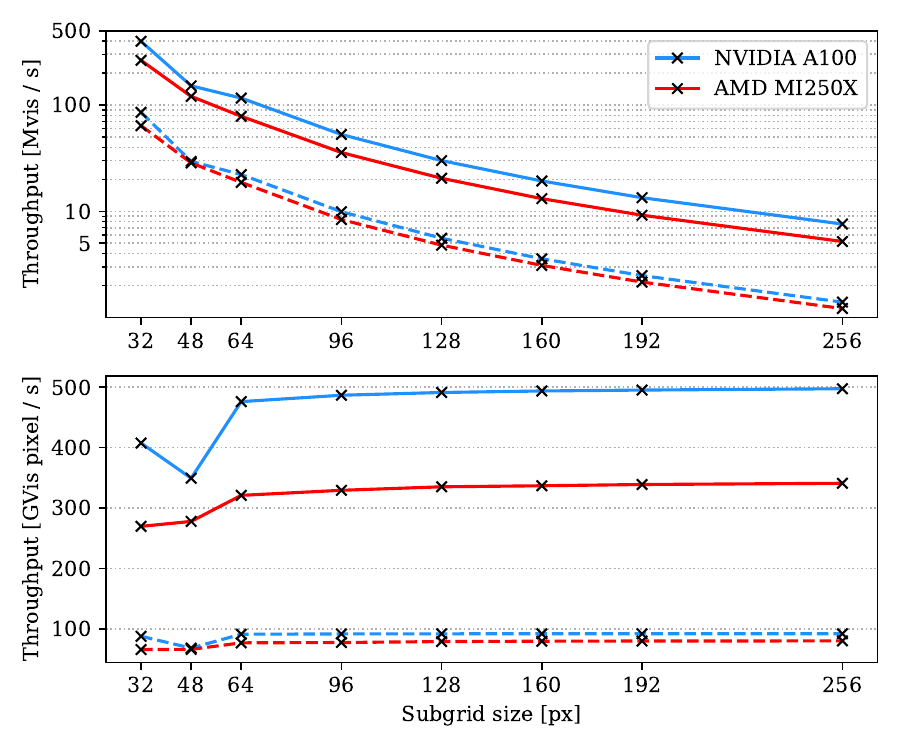}
        \caption{Degridder}
    \end{subfigure}
    \caption{Kernel throughput as a function of subgrid size for both \textsc{amd} \textsc{mi250x} and \textsc{nvidia} \textsc{a100}. Solid and dashed lines indicate single and double precision, respectively.}
    \label{fig:kernelsize}
\end{figure*}

\autoref{fig:kernelsize} presents the visibility throughput for each of the (de)gridder kernels. In the top row, we present the throughput as a function of subgrid size, $L$ (recall that both kernels have complexity $\mathcal{O}(N L^2)$); in the lower row we present the combined (visibility $\times$ pixel) throughput, which we would expect to be constant.

With a subgrid size of 32 pixels, the \textsc{a100} achieves a throughput of 480 and 400 million visibilities per second, for the gridder and degridder kernels, respectively. The \textsc{mi250x} is about a factor of 1.1 slower for the gridder, at 430 million visibilities per second, and a factor of 1.5 slower for the degridder, at 260 million visibilities per second. To put this in perspective, \citet{Veenboer2020} estimate a \textsc{ska} Low data rate of \SI{6.3}{\giga vis \per \second} for imaging applications, which factors in an estimated 10 inversion and prediction cycles as part of cleaning. At these rates, Pigi would require approximately 15 \textsc{a100} \textsc{gpu} devices for gridding and degridding to keep pace, which is well within their projected processing capabilities. Higher resolution subgrids would require more resources: with a subgrid size of 160 pixels, for example, the compute requirements would increase by a factor of 25.

All devices and configurations show nearly constant performance per (visibility $\times$ pixel), with the exception of a small dip for subgrid sizes of 48 pixels. The dip points to the importance of matching subgrid sizes to the launch configuration of kernels. On the \textsc{a100}, we use 128 threads per block for the gridder kernel, and further use a thread coarsening factor of 4. For all kernel sizes sampled here, with the exception of $L=48$ pixels, $(4 \times 128)$ divides $L^2$ without remainder. In the case of a 48 pixel subgrid, however, there is a remainder, and these threads are forced to `come along for the ride', tying up \textsc{gpu} resources even if their results are not used. On the \textsc{mi250x}, on the other hand, we use 64 threads per block for the gridder kernel and as this \textit{does} evenly divide $48^2$ we observe no degradation in performance.

When moving from single to double precision, both kernels display a decrease in performance by a factor of about 4.5. Besides the additional precision required of the floating point pipeline, there are a number of other effects that precision has on the (de)gridder kernels, namely: the register count of the kernels increases and as a result occupancy declines; shared memory usage doubles as does its throughput; and while global memory reads remain unchanged, global writes are now twice as large. The \textsc{a100} has a single to double performance ratio of 1:2 which accounts for some of its reduced performance, whilst the rest is attributable to the double precision \texttt{sincos()} function, which, unlike its single precision counterpart, operates at a lower error tolerance. The \textsc{mi250x}, on the other hand, has a 1:1 ratio of single to double precision computation. As with the \textsc{a100}, part of its reduced performance is attributable to the more precise trigonometric functions, contributing to about 75\% of its performance degradation. The source of the remainder of the degradation on the \textsc{mi250x} remains unclear, but decreased \textsc{gpu} occupancy, which halves as a result of increased register pressure, is likely a contributing factor.

\subsection{Subgrid occupancy}
\label{sec:subgridcount}

\begin{figure*}
    \begin{subfigure}{0.49\textwidth}
        \centering
        \includegraphics[width=1\linewidth,clip,trim={0 0.3cm 0 -0.3cm}]{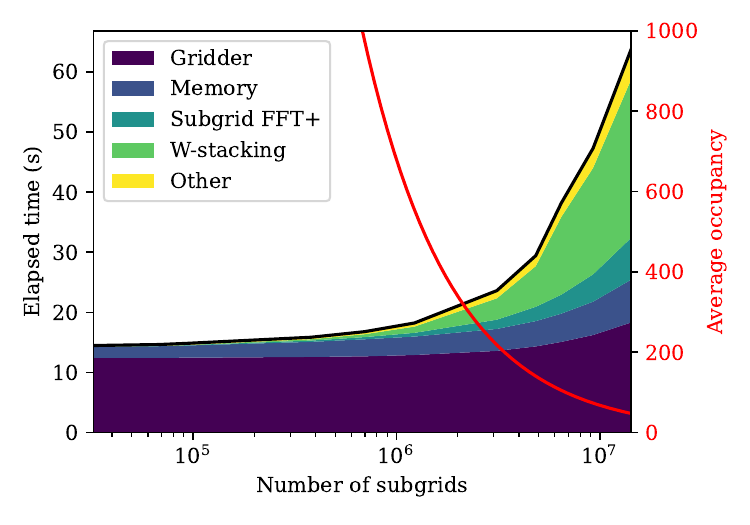}
        \caption{Imaging}
    \end{subfigure}
    \begin{subfigure}{0.49\textwidth}
        \centering
        \includegraphics[width=1\linewidth,clip,trim={0 0.3cm 0 -0.3cm}]{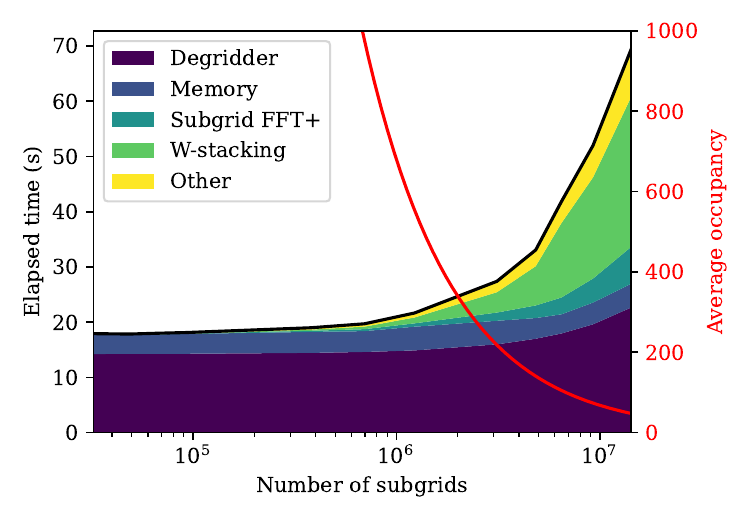}
        \caption{Prediction}
    \end{subfigure}
    \caption{Runtime as a function of subgrid count, including a breakdown of subroutine timings.}
    \label{fig:workunits}
\end{figure*}

We define subgrid occupancy as the number of visibilities assigned to a subgrid. Subgrid occupancy varies depending on the overall image size and the pixel scale. We stated earlier that the complexity of the (de)gridder kernels is $\mathcal{O}(N L^2)$, however, for configurations that significantly suppress subgrid occupancy we observe a deviation from this complexity. Moreover, low occupancy has significant performance impact on the other aspects of the imaging and prediction routines.

In \autoref{fig:workunits}, we show the effect of the total number of subgrids on imaging and prediction duration for a test data set.\footnote{The test data set comprised of $\sim670$ million visibilities, and was imaged at $2000 \times 2000$ pixels using a subgrid size of $96 \times 96$ pixels}. In red, we plot the average visibility occupancy per subgrid which is inversely proportional to the subgrid count. Note that these times are for the full inversion and prediction routines, of which the (de)gridder kernels are just one component. We show the runtime breakdown, delineating the contribution from the (de)gridder kernels, memory allocation and transfer overheads, subgrid processing including \textsc{fft}, and $w$-stacking. 

In both imaging and prediction, we see that the (de)gridding kernels dominate the overall runtime for configurations with a subgrid occupancy greater than about 200. We also see that these kernels are highly resilient to changes in the subgrid count: despite the subgrid count changing by orders of magnitude, the kernel runtimes remain steady and only begin to trend up slightly when the subgrid occupancy becomes less than 200. To explain this uptick, recall that both kernels are distributed over the \textsc{gpu} y-dimension one block per subgrid, and in the x-dimension one thread per subgrid pixel. There are fixed costs per block, such as the computation of $A A^\dagger$ for each subgrid cell, as well as memory reads that are independent of the visibility count. In cases with low subgrid occupancy, these fixed costs are no longer amortised by the computation of the visibilities and instead start to dominate the kernel runtime.

The greatest contribution to the increase in imaging and prediction runtimes are the $w$-stacking routines. As discussed in \autoref{sec:wstacking}, $w$-stacking is a fixed cost that depends primarily on the image size. This fixed cost, however, must be paid in full for each data batch sent to the \textsc{gpu} for (de)gridding. The rising subgrid count means that the memory required to store the subgrids themselves competes with the memory required to store the visibility data. Pigi mitigates this memory pressure by reducing the batch size. As a result, the $w$-stacking routines are forced to process an increasingly smaller number of visibilties per batch, but must nonetheless pay the full cost of shifting the mastergrid through each $w$-layer.

Future work could address the computation involved in $w$-stacking. One straightforward mitigation is to reduce the size of mastergrid to the minimum size required to critically sample the sky. Another candidate is to implement a $w$-tiling scheme as now used in the original \textsc{idg} implementation which takes advantage of the fact that, for an approximately colinear array, only a slice of $(u,v,w)$ space tends to be sampled \citep{Veenboer2020}.

\subsection{Image size}

\begin{figure*}
    \begin{subfigure}{0.49\textwidth}
        \centering
        \includegraphics[width=1\linewidth,clip,trim={0 0.4cm 0 0}]{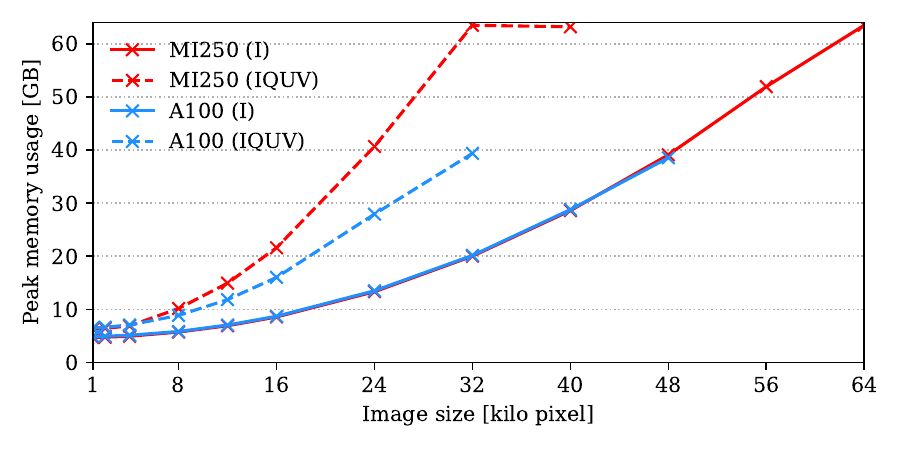}
        \caption{\textsc{gpu} memory usage}
        \label{fig:imagesize-memory}
    \end{subfigure}
    \begin{subfigure}{0.49\textwidth}
        \centering
        \includegraphics[width=1\linewidth,clip,trim={0 0.4cm 0 -1cm}]{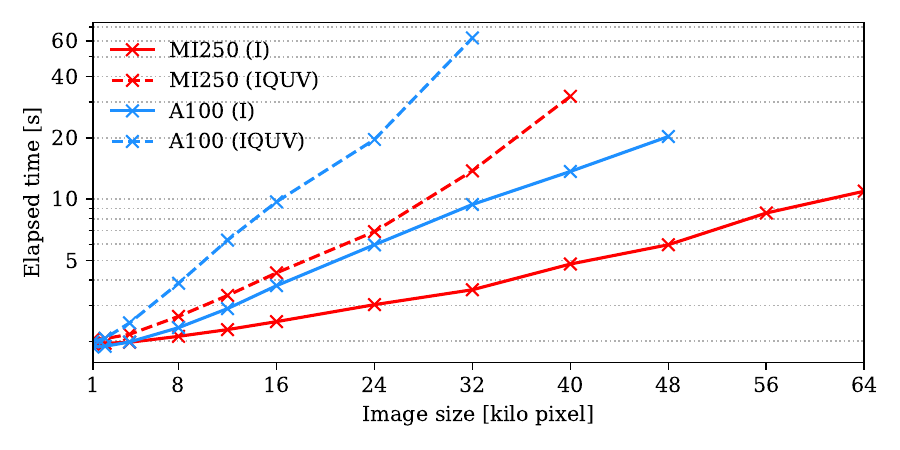}
        \caption{Runtime duration}
        \label{fig:imagesize-time}
    \end{subfigure}
    \caption{The effect of output image size on both memory usage and runtime when performing imaging. Solid lines indicate a single polarization output (Stokes I) and dashed lines indicate four polarisation outputs (Stokes I, Q, U and V).}
    \label{fig:imagesize}
\end{figure*}

Widefield interferometers with increasingly long baselines demand ever larger image sizes spanning tens of thousands of pixels, and it is important that Pigi can handle these requirements.

For Pigi, the primary limitation on image size is memory. In its current design, the entirety of the master grid must fit into \textsc{gpu} device memory. This is a complex valued array requiring 2 floating point values per pixel for Stokes I imaging, and 8 floating point values per pixel for combined Stokes IQUV imaging. This grid must remain resident in device memory as it is frequently modified by the splitter and adder kernels, and during $w$-stacking the entire array will be Fourier transformed multiple times for each $w$-layer. Additionally, the \textsc{fft} plan for this array, including its workspace, must also be resident in device memory. For large images, these two allocations make up the majority of the \textsc{gpu} memory usage.

\autoref{fig:imagesize-memory} shows the peak memory usage for imaging at various sizes. The \textsc{a100} has \SI{40}{\giga \byte} of device memory and is capable of Stokes I imaging up to 48k $\times$ 48k pixels, and for combined Stokes IQUV it is capable of imaging up to 32k $\times$ 32k pixels. The \textsc{mi250x} has \SI{64}{\giga \byte} device memory per logical core and is capable of Stokes I imaging up to 64k $\times$ 64k pixels, and combined Stokes IQUV imaging of up to 40k $\times$ 40k pixels. The size limits are identical in the case of prediction.

Image size also has an effect on runtime duration, as shown in \autoref{fig:imagesize-time}, which owes principally to the multiple \textsc{fft} operations performed over the master grid during $w$-stacking. This is compounded by the fact that, for larger fields of view, the value of $w_\text{max}$ decreases leading to a corresponding increase in the number of $w$-layers. The runtime of the gridding and degridding kernels themselves are unaffected by image size.

\subsection{Scalability}
\label{sec:scalability}

\begin{figure}
    \centering
    \includegraphics[width=1\linewidth]{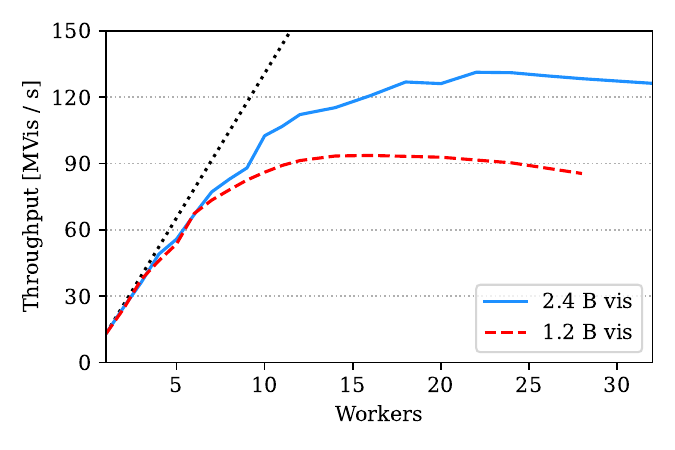}
    \caption{The imaging performance of Pigi as a function of node count. Throughput initially increases linearly with worker count, but quickly falls off due to fixed costs for each worker. Larger datasets scale better than smaller ones, as demonstrated by the 2.4 versus 1.2 billion visibility datasets measured here.}
    \label{fig:nodes}
\end{figure}

As discussed in \autoref{sec:mpi}, Pigi can distribute visibility ownership across \textsc{mpi} ranks and scale to multiple \textsc{gpu}s across multiple nodes. One motivation for this is to distribute a data set across multiple nodes so as to allow for the visibility data to reside in-memory and avoid expensive disk read/write operations. Here we discuss a second motivation, however, which is to accelerate imaging and prediction by utilising more compute resources.

In \autoref{fig:nodes} we show the visibility throughput as a function of worker count, where each worker is allocated an exclusive \textsc{mi250x} logical core. These results are for imaging a 9k $\times$ 9k pixel image, with a padding factor of 1.5, and a large $160 \times 160$ pixel subgrid size. The solid and dashed lines indicate data sets of differing sizes, with 2.4 billion visibilities and 1.2 billion visibilities, respectively. For both data set sizes, as the worker count is increased we observe at first an approximately linear increase in throughput, and then from about 4 workers this quite rapidly diverges from the ideal linear improvement. By 6 workers, for example, the imaging throughput for the 2.4 billion visibility data set is already at just 86\% of the ideal throughput. For larger node counts, the throughput plateaus. For inversion, the 2.4 Bvis data set peaks at \SI{131}{\mega vis \per \second} compared to \SI{94}{\mega vis \per \second} for the smaller 1.2 Bvis data set.

The reason for this poor parallelisation owes to duplication of work amongst workers. The two main components of work for each worker are the (de)gridding kernels and $w$-stacking. The kernels scale linearly with the data volume, and distributing the data volume across multiple nodes is an effective way to accelerate the kernels. $w$-stacking, on the other hand, is duplicated on each node. Each worker will still have to process approximately the same number of $w$-layers, and this work forms a fixed cost per node irrespective of the data volume. This is to say: increasing the worker count will result in runtime improvements only as long as the (de)gridding kernel runtimes dominate per worker. This condition is met when the volume of data per node is large, and this explains why the 2.4 Bvis data set achieves higher throughput.

This reinforces the importance in future work of minimising the $w$-stacking work, as we have already identified in \autoref{sec:subgridcount}.

\subsection{Comparison to \textsc{wsclean}}

\begin{figure}
    \centering
    \includegraphics[width=1\linewidth]{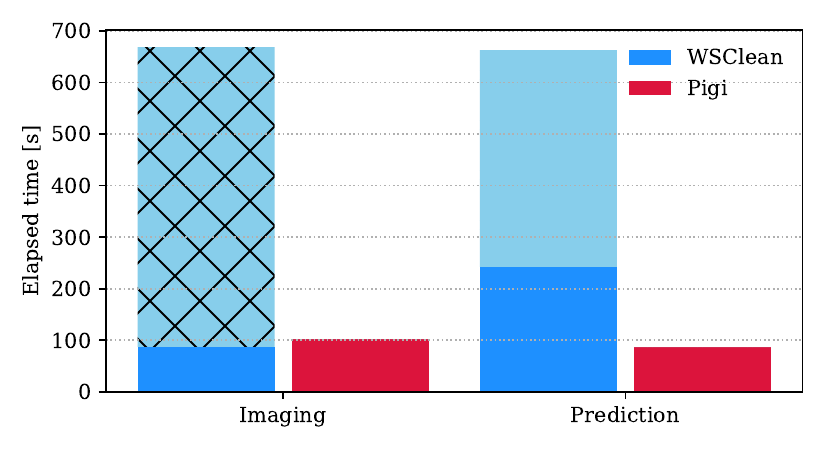}
    \caption{Comparison of runtimes for each of Pigi and \textsc{wsclean} using the \textsc{idg} backend performing a series of 10 inversion and prediction cycles, as benchmarked on an \textsc{nvidia} \textsc{a100}. The hatched segments represent additional runtime reported by \textsc{wsclean} for each of imaging and prediction, dominated by disk read/write operations.}
    \label{fig:wsclean}
\end{figure}

\textsc{wsclean} \citep{Offringa2014} is presently one of the most commonly used imagers in radio astronomy. It can be configured to use multiple backends during inversion and prediction, including the original \textsc{idg} implementation.\footnote{Available at \url{https://idg.readthedocs.io/}.}

Here we compare the runtimes of Pigi and \textsc{wsclean} (with \textsc{idg} backend) when performing 10 rounds of inversion/prediction on a standard \textsc{mwa} snapshot, plus one additional inversion to construct the \textsc{psf}. These runtimes are measured on the \textsc{a100} only, since the \textsc{idg} implementation does not support \textsc{roc}m systems. In this comparison, the configuration parameters for each of Pigi and \textsc{wsclean} are set as similarly as possible: the image size is 8k $\times$ 8k pixels with a padding factor of 1.5, a subgrid size of 128 pixels, full \textsc{mwa} beam correction, and a 30 second maximum beam duration. Whilst we have endeavoured to make this comparison fair, it is possible that differences exist in configuration as well as runtime reporting.

In \autoref{fig:wsclean} we show the total runtimes for both imaging and prediction routines. The \textsc{wsclean} runtimes need some clarification. Its \textsc{idg} implementation reports its own subroutine timings for ``gridding'' and ``degridding'', which we show in the solid colours. The hatched timings show additional runtime reported by \textsc{wsclean} during each iteration of imaging and prediction. This additional time is primarily attributable to data load/store operations. \textsc{wsclean} does not store data in memory even if capacity allows, and so each iteration of imaging and prediction requires reading and writing the visibility data to disk, as well as considerable in-memory data movement.

According to the total reported runtimes, \textsc{wsclean} with \textsc{idg} imaging is about a factor of 6.5 and 7.7 times slower than Pigi, for inversion and prediction respectively. If we ignore the hatched regions, we see that \textsc{wsclean}'s \textsc{idg} implementation is slightly faster for imaging by a factor of 1.16, while it is 2.81 times slower than Pigi for prediction.

Whilst these results are not rigorous, they show that the performance of Pigi is certainly competitive with the original implementation of \textsc{idg} whilst offering broader compatibility.

\section{An \textsc{mwa} case study}

\begin{figure}
    \centering

    \includegraphics[width=1\linewidth]{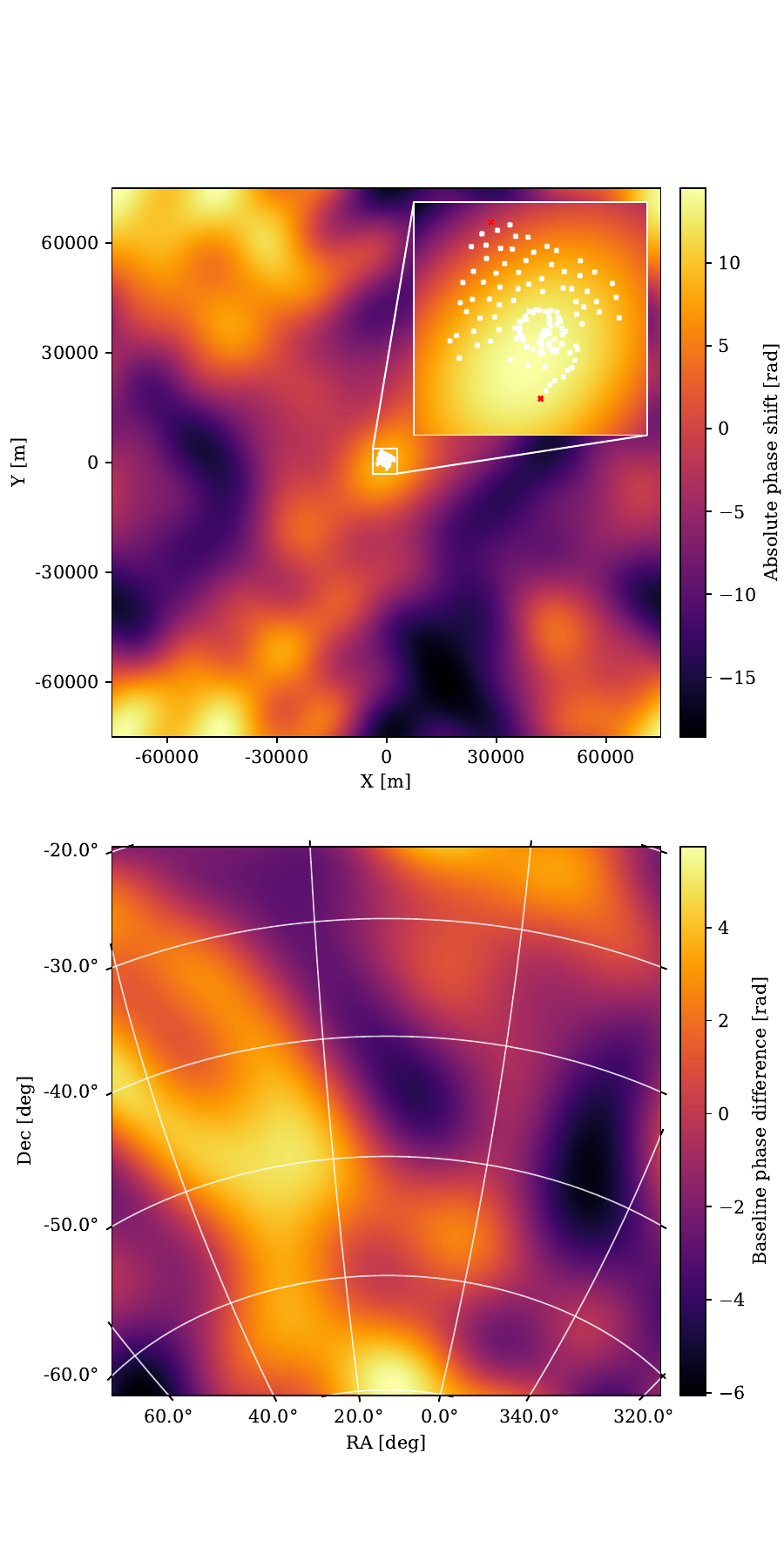}
    \caption{The ionospheric model used in this \textsc{mwa} simulation. \textit{Top:} The ionosphere is modelled as a thin-film phase layer at \SI{100}{\kilo \meter} altitude above the \textsc{mwa} array. As an example, we show the projection of the \textsc{mwa} antennas onto this film for a source at zenith. The  phase difference across the projection gives rise to ionospheric distortions. \textit{Bottom:} The phase difference for the one of the longest baseline pairs in the \textsc{mwa} (highlighted in red). In Pigi, this will be sampled at the resolution of the subgrid and used as a phase correction map.}
    \label{fig:phasescreen}
\end{figure}

\begin{figure*}
    \begin{subfigure}{0.49\linewidth}
        \centering
        \includegraphics[width=1\linewidth,clip,trim={3cm 5cm 3cm 3cm}]{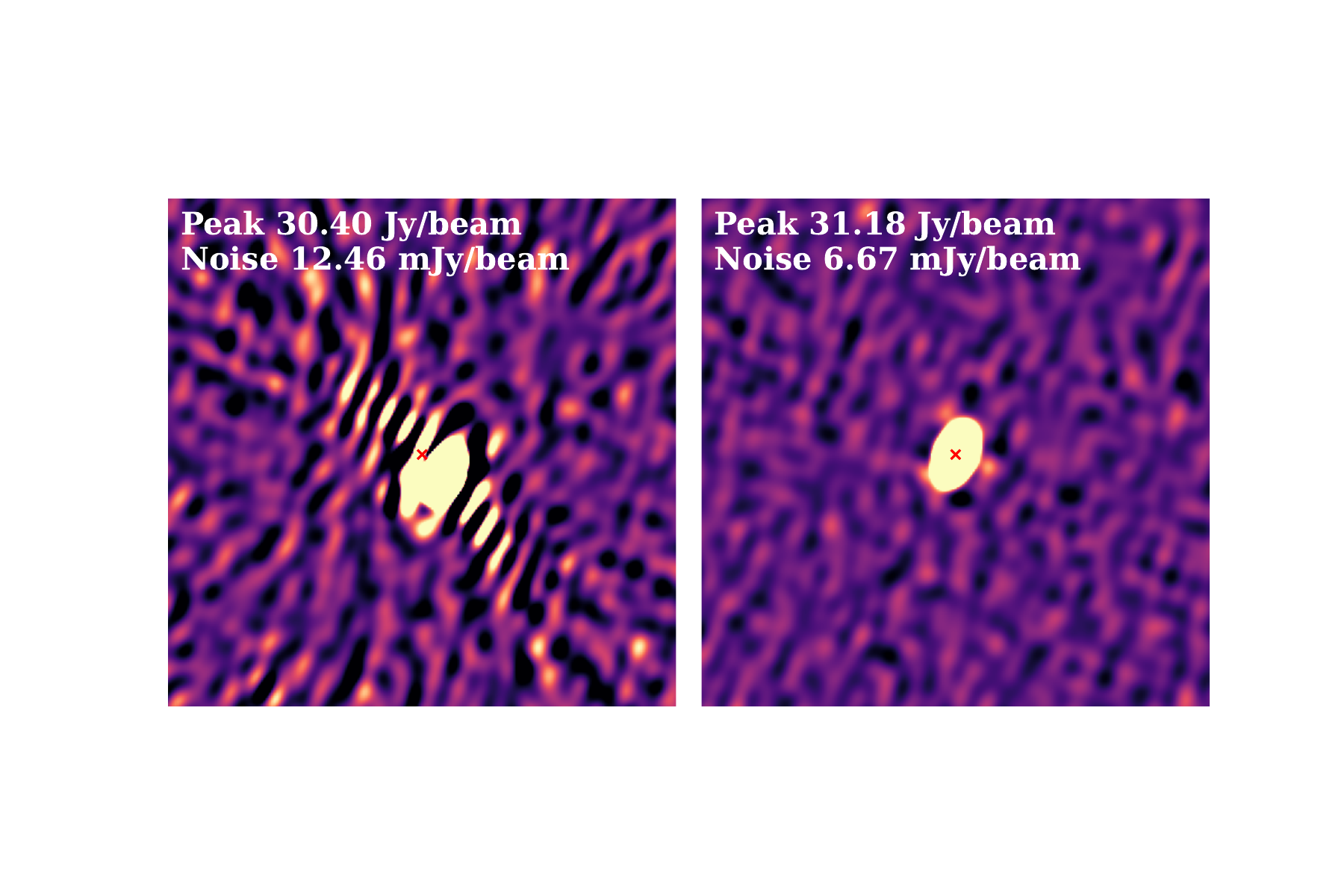}
        \caption{}
        \label{fig:ionosphere-beforeafter-a}
    \end{subfigure}\begin{subfigure}{0.49\linewidth}
        \centering
        \includegraphics[width=1\linewidth,clip,trim={3cm 5cm 3cm 3cm}]{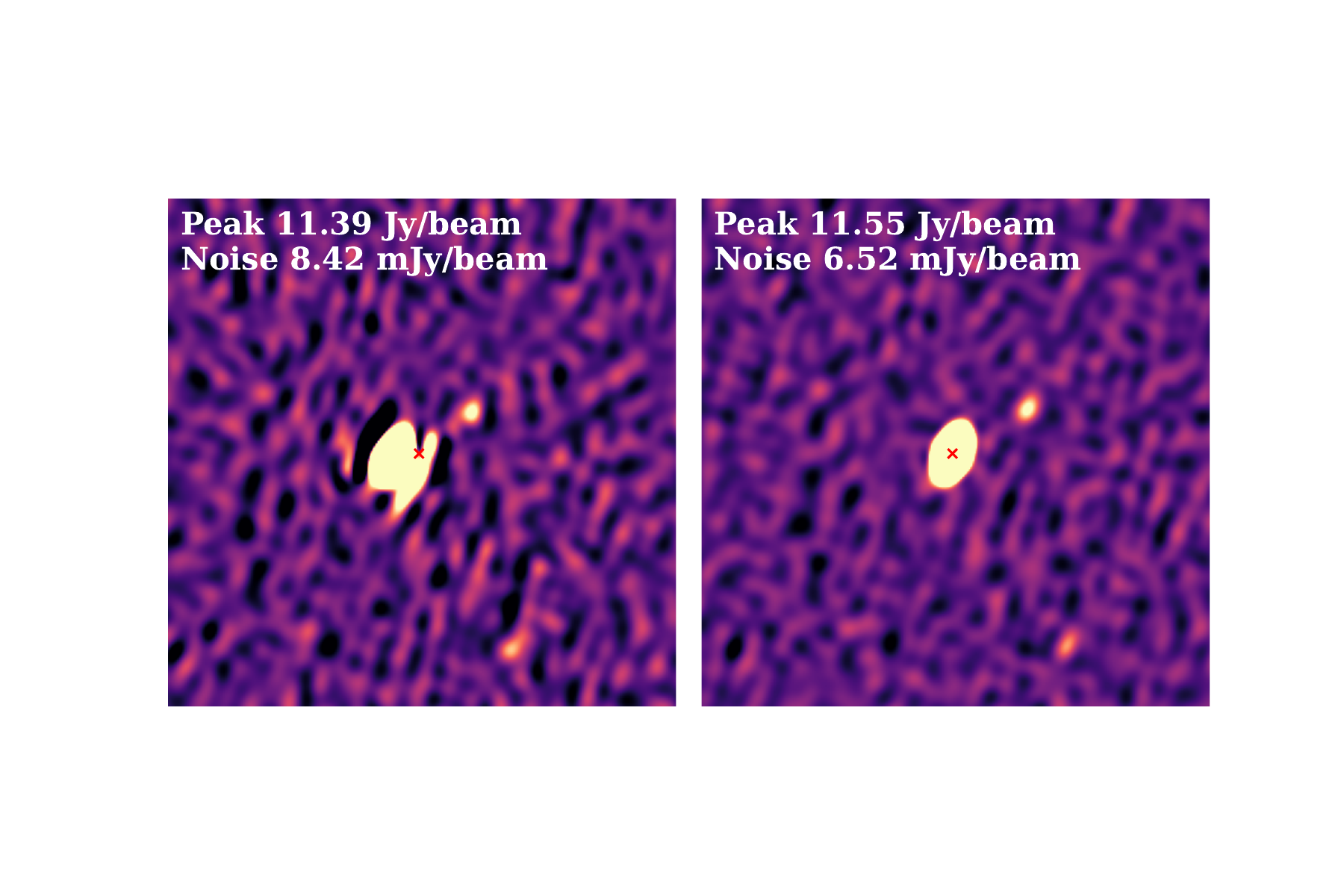}
        \caption{}
    \end{subfigure}
    \begin{subfigure}{0.49\linewidth}
        \centering
        \includegraphics[width=1\linewidth,clip,trim={3cm 5cm 3cm 3cm}]{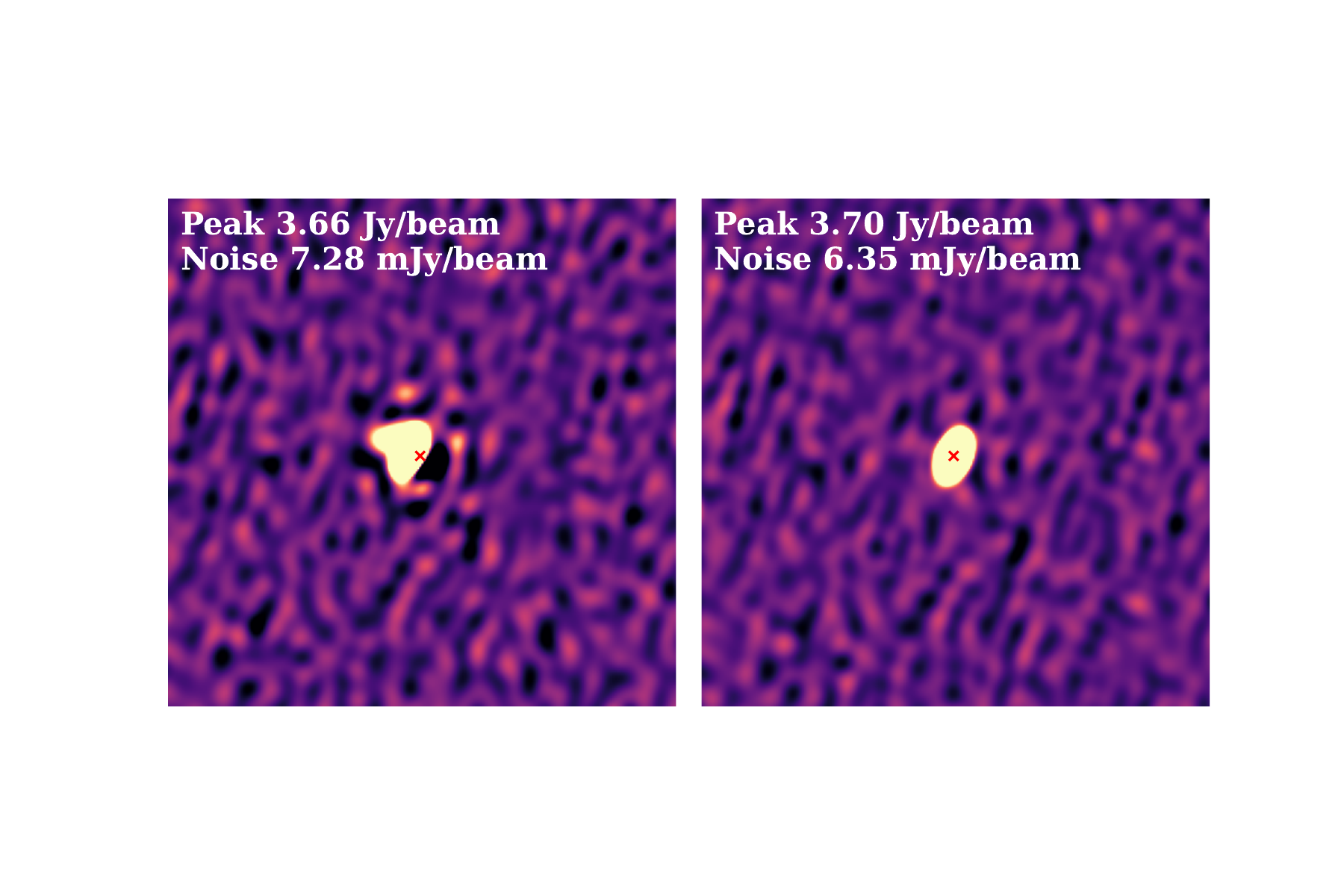}
        \caption{}
    \end{subfigure}\begin{subfigure}{0.49\linewidth}
        \centering
        \includegraphics[width=1\linewidth,clip,trim={3cm 5cm 3cm 3cm}]{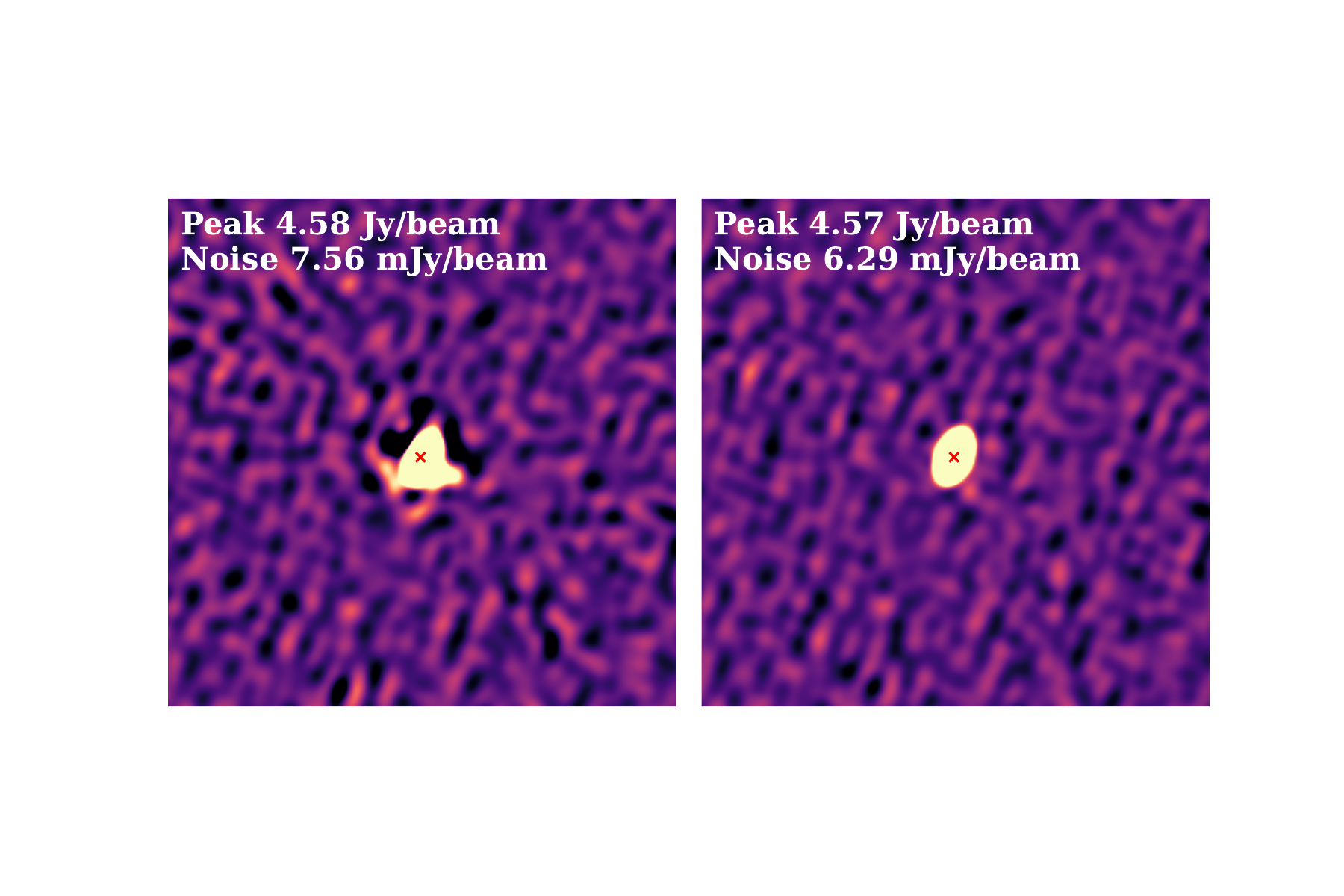}
        \caption{}
    \end{subfigure}
    \caption{Examples of bright sources within a single simulated \textsc{mwa} snapshot, imaged at 16k $\times$ 16k pixels. In each image, the true source location is indicated by the red cross. Noise properties are measured in the region local to the source. \textit{Left:} The output of standard imaging with no $A$-term corrections show the effect of ionospheric distortion on the source. \textit{Right:} The source imaged with ionospheric corrections applied, using a $160 \times 160$ pixel $A$-term kernel.}
    \label{fig:ionosphere-beforeafter}
\end{figure*}   

\begin{figure*}

    \centering
    \includegraphics[width=1\linewidth,clip,trim={3cm 0 1cm 0}]{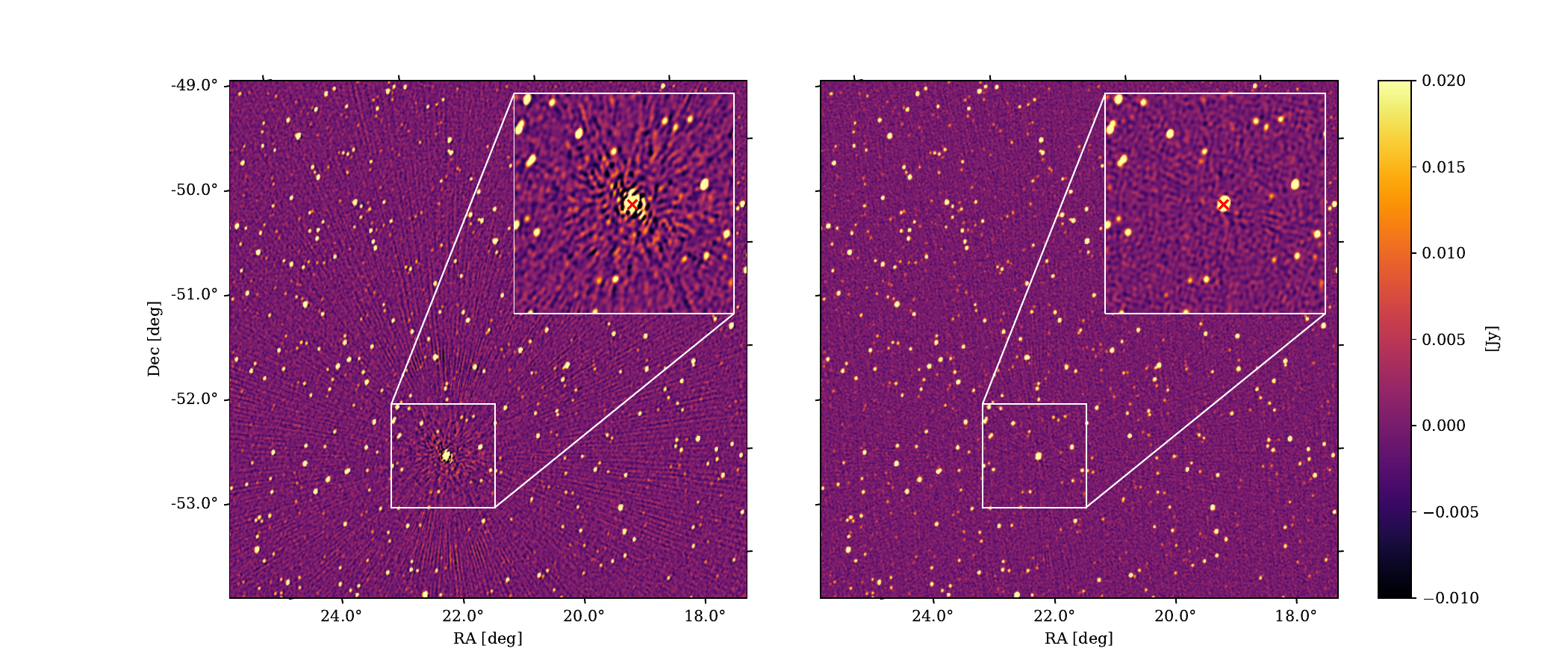}

    \caption{A subsection of the full 32 minute \textsc{mwa} observation, showing a $2000 \times 2000$ pixel subregion drawn from the full $160000 \times 16000$ pixel image. The inset shows the brightest source ($\sim$\SI{31}{\jansky}). \textit{Left:} The stacked image, with each snapshot cleaned down to approximately \SI{24}{\milli \jansky} and warped so as to correct source positions. \textit{Right:} The Pigi image, produced by jointly cleaning all snapshots down to approximately \SI{6}{\milli \jansky}.}
    \label{fig:pipelinecomparison}
\end{figure*}

The \textsc{mwa} has a time-dependent primary beam and suffers from significant ionospheric distortions as a result of its wide field of view and low frequency band. It is an ideal candidate for imaging with $A$-term corrections. In this section we demonstrate the effectiveness of Pigi in applying beam and ionospheric corrections, and compare its output to that of the leading \textsc{mwa} imaging workflow.

\subsection{Stacking and image-domain\\
corrections}

At present, the typical \textsc{mwa} continuum imaging workflow, as exemplified by the GLEAM-X pipeline\footnote{See \url{https://github.com/GLEAM-X/GLEAM-X-pipeline}.}, attempts to mitigate time-dependent effects by imaging short duration `snapshots', typically 2 minutes in length. Over this duration, $A$-terms are assumed to be approximately constant. In the course of a single snapshot, the primary beam is kept constant with regards to the local coordinate system (but has a drift with respect to the celestial frame), and ionospheric variations are meanwhile assumed to operate on time-scales greater than 2 minutes. Each snapshot is imaged in isolation and cleaned down some noise threshold. To produce a full integration, $A$-term corrections are applied in the image domain for each snapshot and then combined by stacking.

In this workflow, $A$-term corrections are limited to simple image-based corrections. Primary-beam correction, for example, is typically performed as a scalar correction by correcting for beam power only. Ionospheric corrections are limited to so-called `first order' corrections, that is, to correcting position offsets of sources. This technique is exemplified by FitsWarp \citep{HurleyWalker2018} which, after cross-matching sources, `warps' a snapshot image so as to shift sources back to their true locations.

There are a number of issues with this workflow that impede deep and highly sensitive observations. Firstly, the primary beam can be considered approximately constant over the duration of a snapshot only near the central lobe; sources on the periphery, where the primary beam changes much more rapidly, experience significant variance even over this interval. Secondly, these same peripheral sources often exhibit significant induced polarisation due to differing sensitivity patterns for each of the \textsc{mwa}'s dipole antenna. This apparent polarisation needs to be corrected during deconvolution or else will appear as an amplitude error in the final image. Thirdly, the deconvolution depth is limited by the noise level in individual snapshots. In long integrations, the stacked image will contain a significant amount of residual `dirty' sources that are well above the final integrated noise level, leading to visible negative bowls around sources as well as additional sidelobe confusion noise throughout the image. Finally, baseline-dependent $A$-term corrections cannot be made which prohibits more complex ionospheric corrections (e.g.\@ lensing, elongation or splitting of sources) or station-dependent beams (e.g.\@ tiles with dead dipoles).

\subsection{Data simulation and\\
imaging configuration}

For this comparison, we use 16 snapshot observations from an observing campaign in 2020, spanning 32 minutes in total. We have simulated the visibility data, whilst retaining the metadata, beam configuration, weights, flags and baseline coverage. We have chosen to simulate the visibilities so as to isolate effects that might otherwise be caused by poor calibration, imperfect beam models, or improperly modelled ionospheric effects. The simulated sky catalogue is based on a modified GLEAM catalogue, augmented with a population of faint, randomly distributed sources that ensure source count completeness down to \SI{0.1}{\milli \jansky} (using a parameterisation of $\nicefrac{\dd{N}}{\dd{S}}$ at \SI{154}{\mega \hertz} by \citet{Franzen2019}). The visibility data is averaged to \SI{4}{\second} and \SI{40}{\milli \hertz} and is supplemented with \SI{80}{\jansky} of complex-valued noise intended to stand in for thermal and other noise sources, and to act as a depth floor for cleaning thresholds.

The ionosphere is modelled using a thin phase screen situated \SI{100}{\kilo \meter} above the \textsc{mwa}, as shown in \autoref{fig:phasescreen}. For any given direction on the sky, each antenna will `look through' this screen and will incur whatever phase delay is associated with that point in the screen. Since each antenna is laterally offset on the ground, their associated pierce point through the screen will be similarly offset, naturally giving rise to baseline-dependent phase delays. This is, however, strictly a toy model: the phase screen itself is randomly generated and used here only for demonstration purposes. This ionospheric model is used to distort the visibilities during visibility simulation, with a unique ionosphere constructed for each snapshot. Additionally, it is used to generate the small, subsampled $A$-term kernels used in Pigi as well as to compute the first-order approximation used in the standard \textsc{mwa} pipeline to correct for position offsets.

We produce images of this field using the following imaging parameters. We image 16k $\times$ 16k pixel images with a scale of \SI{9}{\arcsecond} and a padding factor of 1.3. The image spans \SI{41}{\degree} which is necessary to deconvolve sources on the edge of the field owing to the expansive \textsc{mwa} primary beam. A subgrid size of 160 pixels is used, and this is thus the resolution of the ionospheric correction map. All images are cleaned to a factor of 3.5 times the noise: in the case of snapshots, this means they are cleaned to approximately \SI{24}{\milli \jansky}; while in the case of Pigi's joint deconvolution, the map is cleaned to about \SI{6}{\milli \jansky}. In both workflows, visibilities are weighted with a Briggs weighting scheme (robust parameter set to 0) however there is nonetheless some difference in weighting. This difference arises because, in the stacking workflow, snapshots are weighted independently and thus the final stacked image will tend to be slightly more naturally weighted.

\subsection{Comparison}

\autoref{fig:ionosphere-beforeafter} shows examples of a number of bright sources contained in a single, 2-minute snapshot. On the left of each subfigure, we show the source imaged with no $A$-term correction. We observe ionospheric distortions that are typical of the `bad' kind of ionospheric effects that one might encounter with the \textsc{mwa}. Consider \autoref{fig:ionosphere-beforeafter-a}, for example, which is the brightest source in the field and exhibits both `ringing' along one axis about its centre as well as radial spokes that span widely across the image. This source also exhibits a position offset of about an arcminute. Elsewhere throughout the field, the other example sources show similar kinds of ionospheric artefacts and position offsets. On the right of each subfigure we show the output of imaging with Pigi's ionospheric corrections. We can observe that the visual artefacts are largely or wholly resolved, source positions are corrected back to their expected locations (indicated by the red cross), and noise in the local region around each source drops considerably.

\autoref{fig:pipelinecomparison} shows the full 32 minute integration using the two different workflows. On the left we show the results of the \textsc{mwa}'s `stacking' workflow, including snapshot warping to correct position offsets. Once again we notice that the stacked image exhibits serious artefacts about its bright sources, although these have averaged down with additional snapshots. As before, radial spokes extend throughout the image. The position offsets of sources, however, have been corrected and they are now well aligned with their true position. As each snapshot was cleaned individually, the stacked \textit{residual} images show a large population of sources that have not been cleaned and which will contribute to an elevated background sidelobe confusion. Additionally, faint sources show characteristic negative lobes from this shallower cleaning depth.

On the right we show the output of the joint deconvolution over all 16 snapshots produced by Pigi, with both primary beam and ionospheric corrections applied. Here we observe that the visual artefacts about bright sources are largely resolved and position offsets are well corrected. There are no visually detectable negative lobes about faint sources and the residual image appears noise-like, both of which result from cleaning a factor of 4 times deeper. Restored point sources appear more compact than in the stacked case, owing partly to lensing effects which have been corrected by Pigi, but also as the restoring beam is fitted to the \textsc{psf} of the full 32 minute integration.

There are, however, still faint radial spokes around the \SI{35}{\jansky} source that are visually perceptible and which indicate some residual error remains. The single biggest cause of this error arises from interpolation across the low-resolution $A$-term kernel. The spatial scale of some ionospheric variations are not well captured by the $160\times160$ pixel subgrid which, when accounting for padding, has a mean resolution of approximately half a degree. One straightforward method of improvement is to simply increase the size of the $A$-term kernel, although this comes at an increased computational cost. An as-yet unexplored alternative is to construct the $A$-term kernel more intelligently. In the present case, the $A$-term kernel was constructed very simply such that each pixel mapped exactly to the ionospheric phase value at its associated coordinate. A better approach might be to set the pixel values such that the \textit{interpolated} kernel provides a best fit to the flux-weighted sources across the field of view. In this way, the kernel would be modified slightly to preferentially favour corrections for the brighter sources.

\section{Conclusion}

$A$-term effects remain an impediment to science cases that require highly accurate and sensitive radio interferometric observations. Whilst algorithms such as AW-projection have existed for some time that allow for properly handing these terms, they have not seen widespread adoption due to their expensive computational requirements. Instead, simpler, ad-hoc alternatives such as snapshot stacking and faceting have seen widespread use in modern imaging pipelines.

Pigi is a new implementation of the \textsc{idg} algorithm that allows for the application of direction-, time- and baseline-dependent $A$-term corrections. We have shown Pigi is capable of dramatically improving observations that have significant beam- and ionospheric-distortions, and crucially Pigi does so with only modest computational requirements. 

Pigi uses \textsc{gpu} accelerators to achieve this high performance, and is optimised for both \textsc{amd} and \textsc{nvidia} platforms. The performance of the core gridder and degridder kernels depends upon the size of the subgrid kernel. For small $L \times L = 32 \times 32$ pixel subgrids, Pigi can reach a throughput of almost half a billion visibilities per second; for larger-sized subgrids, this performance drops at a rate of $\nicefrac{1}{L^2}$. We have also shown that there remains scope for performance improvements, specifically regarding excessive duplication of work during $w$-stacking, and resolving this will be important to ensuring Pigi scales well in \textsc{hpc} multi-node environments.

Pigi is highly accurate. In the case of spatially-uniform $A$-terms, the accuracy of Pigi is dominated by trigonometric and floating point errors, and can achieve a signal to noise ratio of around \SI{100}{\decibel} when using double precision. In the case of spatially-varying $A$-terms, however, the inaccuracy is dominated by the interpolation of the subgrid. Interpolation across low-resolution kernels can lead to errors in cases where those $A$-terms are rapidly changing or have fine spatial details. To mitigate this, we have shown that increasing the subgrid size---at the expense of a corresponding performance penalty---can have a marked improvement on interpolation accuracy.

For Pigi to be useful with real-world data, it will be necessary to ensure suitable $A$-terms can be generated, and in particular to be able to construct phase screens that can reliably correct for ionospheric distortions. This latter aspect is a missing piece of the puzzle that we believe is essential to making Pigi and, more broadly \textsc{idg}, a viable alternative to the status quo, and represents the next step in our work.

\subsection*{Acknowledgements}

The authors thank the Pawsey Supercomputing Centre for their generous funding of this project as part of the \textsc{pacer} program.

\bibliographystyle{mnras}
\bibliography{refs}

\begin{thebibliography}{}
\makeatletter
\relax
\def\mn@urlcharsother{\let\do\@makeother \do\$\do\&\do\#\do\^\do\_\do\%\do\~}
\def\mn@doi{\begingroup\mn@urlcharsother \@ifnextchar [ {\mn@doi@} {\mn@doi@[]}}
\def\mn@doi@[#1]#2{\def\@tempa{#1}\ifx\@tempa\@empty \href {http://dx.doi.org/#2} {doi:#2}\else \href {http://dx.doi.org/#2} {#1}\fi \endgroup}
\def\mn@eprint#1#2{\mn@eprint@#1:#2::\@nil}
\def\mn@eprint@arXiv#1{\href {http://arxiv.org/abs/#1} {{\tt arXiv:#1}}}
\def\mn@eprint@dblp#1{\href {http://dblp.uni-trier.de/rec/bibtex/#1.xml} {dblp:#1}}
\def\mn@eprint@#1:#2:#3:#4\@nil{\def\@tempa {#1}\def\@tempb {#2}\def\@tempc {#3}\ifx \@tempc \@empty \let \@tempc \@tempb \let \@tempb \@tempa \fi \ifx \@tempb \@empty \def\@tempb {arXiv}\fi \@ifundefined {mn@eprint@\@tempb}{\@tempb:\@tempc}{\expandafter \expandafter \csname mn@eprint@\@tempb\endcsname \expandafter{\@tempc}}}

\bibitem[\protect\citeauthoryear{{Bhatnagar}, {Cornwell}, {Golap}  \& {Uson}}{{Bhatnagar} et~al.}{2008}]{Bhatnagar2008}
{Bhatnagar} S.,  {Cornwell} T.~J.,  {Golap} K.,   {Uson} J.~M.,  2008, \mn@doi [\aap] {10.1051/0004-6361:20079284}, \href {https://ui.adsabs.harvard.edu/abs/2008A&A...487..419B} {487, 419}

\bibitem[\protect\citeauthoryear{{Brouw}}{{Brouw}}{1975}]{Brouw2975}
{Brouw} W.~N.,  1975, \mn@doi [Methods in Computational Physics] {10.1016/B978-0-12-460814-6.50008-5}, \href {https://ui.adsabs.harvard.edu/abs/1975MComP..14..131B} {14, 131}

\bibitem[\protect\citeauthoryear{{Cornwell}, {Golap}  \& {Bhatnagar}}{{Cornwell} et~al.}{2008}]{Cornwell2008}
{Cornwell} T.~J.,  {Golap} K.,   {Bhatnagar} S.,  2008, \mn@doi [IEEE Journal of Selected Topics in Signal Processing] {10.1109/JSTSP.2008.2005290}, \href {https://ui.adsabs.harvard.edu/abs/2008ISTSP...2..647C} {2, 647}

\bibitem[\protect\citeauthoryear{{Franzen}, {Vernstrom}, {Jackson}, {Hurley-Walker}, {Ekers}, {Heald}, {Seymour}  \& {White}}{{Franzen} et~al.}{2019}]{Franzen2019}
{Franzen} T.~M.~O.,  {Vernstrom} T.,  {Jackson} C.~A.,  {Hurley-Walker} N.,  {Ekers} R.~D.,  {Heald} G.,  {Seymour} N.,   {White} S.~V.,  2019, \mn@doi [\pasa] {10.1017/pasa.2018.52}, \href {https://ui.adsabs.harvard.edu/abs/2019PASA...36....4F} {36, e004}

\bibitem[\protect\citeauthoryear{{Hamaker}}{{Hamaker}}{2000}]{Hamaker2000}
{Hamaker} J.~P.,  2000, \mn@doi [\aaps] {10.1051/aas:2000337}, \href {https://ui.adsabs.harvard.edu/abs/2000A&AS..143..515H} {143, 515}

\bibitem[\protect\citeauthoryear{{Hurley-Walker} \& {Hancock}}{{Hurley-Walker} \& {Hancock}}{2018}]{HurleyWalker2018}
{Hurley-Walker} N.,  {Hancock} P.~J.,  2018, \mn@doi [Astronomy and Computing] {10.1016/j.ascom.2018.08.006}, \href {https://ui.adsabs.harvard.edu/abs/2018A&C....25...94H} {25, 94}

\bibitem[\protect\citeauthoryear{{Oak Ridge National Laboratory}}{{Oak Ridge National Laboratory}}{2024}]{Frontier}
{Oak Ridge National Laboratory} 2024, {Frontier}, \url{https://www.olcf.ornl.gov/frontier/}

\bibitem[\protect\citeauthoryear{{Offringa} \& {Smirnov}}{{Offringa} \& {Smirnov}}{2017}]{Offringa2017}
{Offringa} A.~R.,  {Smirnov} O.,  2017, \mn@doi [\mnras] {10.1093/mnras/stx1547}, \href {https://ui.adsabs.harvard.edu/abs/2017MNRAS.471..301O} {471, 301}

\bibitem[\protect\citeauthoryear{{Offringa} et~al.,}{{Offringa} et~al.}{2014}]{Offringa2014}
{Offringa} A.~R.,  et~al., 2014, \mn@doi [\mnras] {10.1093/mnras/stu1368}, \href {https://ui.adsabs.harvard.edu/abs/2014MNRAS.444..606O} {444, 606}

\bibitem[\protect\citeauthoryear{{Offringa}, {Mertens}, {van der Tol}, {Veenboer}, {Gehlot}, {Koopmans}  \& {Mevius}}{{Offringa} et~al.}{2019}]{Offringa2019}
{Offringa} A.~R.,  {Mertens} F.,  {van der Tol} S.,  {Veenboer} B.,  {Gehlot} B.~K.,  {Koopmans} L.~V.~E.,   {Mevius} M.,  2019, \mn@doi [\aap] {10.1051/0004-6361/201935722}, \href {https://ui.adsabs.harvard.edu/abs/2019A&A...631A..12O} {631, A12}

\bibitem[\protect\citeauthoryear{{Pawsey Supercomputing Center}}{{Pawsey Supercomputing Center}}{2024}]{Pawsey}
{Pawsey Supercomputing Center} 2024, {Setonix}, \url{https://pawsey.org.au/systems/setonix/}

\bibitem[\protect\citeauthoryear{{Schwab}}{{Schwab}}{1984}]{Schwab1984}
{Schwab} F.~R.,  1984, \mn@doi [\aj] {10.1086/113605}, \href {https://ui.adsabs.harvard.edu/abs/1984AJ.....89.1076S} {89, 1076}

\bibitem[\protect\citeauthoryear{{Smirnov}}{{Smirnov}}{2011}]{Smirnov2011}
{Smirnov} O.~M.,  2011, \mn@doi [\aap] {10.1051/0004-6361/201116764}, \href {https://ui.adsabs.harvard.edu/abs/2011A&A...531A.159S} {531, A159}

\bibitem[\protect\citeauthoryear{{Thompson}, {Moran}  \& {Swenson}}{{Thompson} et~al.}{2017}]{Thompson2017}
{Thompson} A.~R.,  {Moran} J.~M.,   {Swenson} George~W. J.,  2017, {Interferometry and Synthesis in Radio Astronomy, 3rd Edition}, \mn@doi{10.1007/978-3-319-44431-4.
}

\bibitem[\protect\citeauthoryear{{Tingay} et~al.,}{{Tingay} et~al.}{2013}]{Tingay2013}
{Tingay} S.~J.,  et~al., 2013, \mn@doi [\pasa] {10.1017/pasa.2012.007}, \href {https://ui.adsabs.harvard.edu/abs/2013PASA...30....7T} {30, e007}

\bibitem[\protect\citeauthoryear{{Veenboer} \& {Romein}}{{Veenboer} \& {Romein}}{2020}]{Veenboer2020}
{Veenboer} B.,  {Romein} J.~W.,  2020, \mn@doi [Astronomy and Computing] {10.1016/j.ascom.2020.100386}, \href {https://ui.adsabs.harvard.edu/abs/2020A&C....3200386V} {32, 100386}

\bibitem[\protect\citeauthoryear{Veenboer, Petschow  \& Romein}{Veenboer et~al.}{2017}]{Veenboer2017}
Veenboer B.,  Petschow M.,   Romein J.~W.,  2017, in 2017 IEEE International Parallel and Distributed Processing Symposium (IPDPS). pp 545--554, \mn@doi{10.1109/IPDPS.2017.68}

\bibitem[\protect\citeauthoryear{{Wayth} et~al.,}{{Wayth} et~al.}{2015}]{Wayth2015}
{Wayth} R.~B.,  et~al., 2015, \mn@doi [\pasa] {10.1017/pasa.2015.26}, \href {https://ui.adsabs.harvard.edu/abs/2015PASA...32...25W} {32, e025}

\bibitem[\protect\citeauthoryear{{Williams} et~al.,}{{Williams} et~al.}{2021}]{Williams2021}
{Williams} W.~L.,  et~al., 2021, \mn@doi [\aap] {10.1051/0004-6361/202141745}, \href {https://ui.adsabs.harvard.edu/abs/2021A&A...655A..40W} {655, A40}

\bibitem[\protect\citeauthoryear{{van Haarlem} et~al.,}{{van Haarlem} et~al.}{2013}]{vanHaarlem2013}
{van Haarlem} M.~P.,  et~al., 2013, \mn@doi [\aap] {10.1051/0004-6361/201220873}, \href {https://ui.adsabs.harvard.edu/abs/2013A&A...556A...2V} {556, A2}

\bibitem[\protect\citeauthoryear{{van der Tol}, {Veenboer}  \& {Offringa}}{{van der Tol} et~al.}{2018}]{VanDerTol2018}
{van der Tol} S.,  {Veenboer} B.,   {Offringa} A.~R.,  2018, \mn@doi [\aap] {10.1051/0004-6361/201832858}, \href {https://ui.adsabs.harvard.edu/abs/2018A&A...616A..27V} {616, A27}

\makeatother
\end{thebibliography}

\end{document}